Tradução de

# TENTAMEN THEORIAE DE FRICTIONE FLVIDORVM*

## Auctore

## L. EVLERO

Tentativa de uma Teoria da Fricção dos Fluidos

Pelo Autor

L. Euler

Traduzido por Sylvio R. Bistafa†

---

SUMÁRIO

Vários experimentos atestam que os fluidos sofrem certa fricção, enquanto são transportados através de canais, e tanto que os Fisiologistas não hesitam em atribuir o calor animal à fricção que afeta o movimento do sangue; mas isto, no entanto, de modo nenhum foi até agora satisfatoriamente provado, contudo é certo que, se a água for desviada para fontes de jatos d'água através de longos canais, o seu movimento é não pouco retardado por causa da fricção, por isso que a altura do jato a tal ponto tanto mais se afasta da altitude da água no reservatório, quanto mais longo for o curso e mais estreitos os próprios canais. Além disso, não há dúvida de que esta fricção siga leis similares, como a fricção dos corpos sólidos, existindo menores partículas do fluido, que escoam junto das laterais dos canais, possam ser consideradas como sólidos, por causa disto a fricção que é observada deva ser sempre estimada proporcional à pressão. Visto que verdadeiramente as partículas dos fluidos são muitíssimo escorregadias, a fricção delas deve ser sem dúvida estimada muito menor, do que nos sólidos, para a qual também isto convém que se concorde, que nos fluidos, somente as partículas consideradas contíguas às laterais dos tubos sofram fricção, enquanto que próximo aos interiores, estas podem fluir livremente, donde é necessário que o efeito desta fricção torne-se muito menor. Mas visto que acontece

---

*Originalmente publicado em *Novi Commentarii academiae scientiarum Petropolitanae* **6**, 1761, pp. 338-388. Índice de Eneström **E260**.

†Escola Politécnica da Universidade de São Paulo, 05508-900, São Paulo, SP, Brasil
e-mail: sbistafa@usp.br

***Opera Omnia*: Series 2, Volume 12, pp. 169 – 214: Reprodução de 1954 da Edição primeira editada por Clifford Truesdell (C.T.). A presente tradução foi feita a partir desta Reprodução. As figuras foram extraídas da Edição primeira.

desta maneira, já que em qualquer secção do tubo as partículas interiores são movimentadas mais rapidamente, que aquelas que estão contíguas às laterais, contanto que a hipótese, para a qual, a teoria do movimento dos fluidos através dos tubos de escoamento é construída, é derrubada, posto que as conclusões imediatamente deduzidas do cálculo se afastam muito da verdade. Pois estas são transmitidas pelo muito Famoso Euler nesta dissertação à respeito da fricção da água através dos tubos que escoam, na mesma o próprio Autor apregoa simples como certa e consentânea à verdade; mas antes existe todo um estudo diante da tentativa de investigação desta fricção na natureza, com a qual quer fornecer uma oportunidade para os que sondam a natureza, tanto através de experimentos, como através de cálculos acurados. Portanto, assume a conhecida hipótese, pela qual, em qualquer parte da secção do tubo, a água é colocada para ser levada à diante com movimento uniforme e ainda estabelece ser submetida à fricção proporcional à pressão da água contra os flancos do tubo, que, no entanto, a partir de certos experimentos grosseiros, considera mil vezes menor que nos sólidos, para a mesma pressão. No entanto, primeiro observa que nos tubos cilíndricos o efeito da fricção é duas vezes menor, do que nos tubos de mesma amplitude, cujas secções são quadradas. Então, verdadeiramente, determina a diminuição do movimento da água através de quaisquer tubos se verticais ou se inclinados em relação ao horizonte e define os limites em todos os casos, tanto no comprimento dos tubos, quanto das estreitezas dos mesmos, onde o movimento da água diante da fricção é contido internamente, de sorte que estes mesmos limites possam ser facilmente explorados através de experimentos, daqui especialmente é obtida uma forma conveniente de completar a teoria da fricção dos fluidos por meio de experimentos.

Além disso, o Autor adapta esta teoria ao curso das correntes até então[1] não pouco imperfeita (pág. 198[2]), onde observa que a água vai se estagnar, a não ser que a declividade da corrente supere certo limite em presença de dada profundidade do leito: de tal maneira que assumida a mesma medida da fricção, a qual alguns grosseiros experimentos pareciam aconselhar, é logo induzida a corrente da qual a profundidade seja 25 pés e a declividade do leito para uma distância de 1000 pés fosse menor que 9 polegadas[3]: mas se a profundidade for 15 pés, é suficiente para o escoamento, diante de uma distância de 1000 pés, que a declividade supere meio pé: daqui também, é considerado que o efeito da fricção é demasiadamente grande.

---

[1]Edição primeira: 'teoria também até então'. Baseado na versão do sumário contido no manuscrito desta edição. A.S.
Nota do Tradutor: presume-se que A.S. seja Andreas Speiser, um dos produtores da Reprodução de 1954.

[2] Nota do Tradutor: Caso IV.

[3] Nota do Tradutor: esta estimativa de declividade está incorreta, se tiver sido baseada na equação do parágrafo 91, pois para a profundidade $g = 25$ pés, esta equação fornece: $z > \frac{30}{4g} = 0{,}3$ pé $= 3{,}6$ polegadas de declividade. Por outro lado, de acordo com esta mesma equação, a menor declividade para que a água escoe $z = 9$ polegadas $=$ ¾ pé, seria para a profundidade de 10 pés, e não de 25 pés.

É anexado o apêndice (pág. 204) acerca das fontes de jatos d'água que para tal, esta teoria foi aplicada e, quanto pode ser feito a partir dela, mostrando claramente, para qual altitude a água para o alto será projetada nas fontes deste tipo, sendo conhecida a elevação do reservatório e não a amplitude tampouco o comprimento. Também aqui é principalmente apropriado de notar como calculamos o efeito desta fricção também dependendo da pressão da atmosfera, de tal modo que o efeito da fricção seja aumentado igualmente com a pressão da atmosfera aumentada: razão pela qual se a água é desviada através de canais para fontes de jatos d'água, se deve considerar este singular fenômeno local, que a água seja projetada mais alto, quanto mais baixo mantenha-se o mercúrio no barômetro.

Finalmente (pag. 207[4]) é juntada uma longa tabela, da qual diante do comprimento e amplitude dos canais, se pode definir facilmente a altura para a qual a água é elevada, para qualquer altitude da água no reservatório, onde em geral será útil ter ele notado, que quanto mais longos e simultaneamente mais estreitos forem os canais, através dos quais a água é derivada, maior será o prejuízo na altura de qualquer fonte.

Portanto, muito orienta as fontes de jatos d'água a serem construídas, que o reservatório seja estabelecido mais próximo e simultaneamente que sejam adotados os tubos mais amplos. Verdadeiramente, além disso, é prontamente advertido, que esta teoria precisa de correção, que dificilmente será posta adiante com sucesso, antes de avaliar os muitos novos experimentos que com muito cuidado tenham sido planejados, que mais adiante é inteiramente conveniente, na qual tanto os Físicos como os Geômetras empreguem as suas forças com todo o estudo.

## I.

Uma brilhante experiência confirma, que a água, enquanto através de canais é movida para frente, não sofre pouca diminuição do seu movimento, sendo que esta é tanto mais notável, quanto mais longínquos tenham sido, e simultaneamente mais estreitos, os canais deste tipo. E também esta diminuição de movimento é percebida em primeiro lugar nas fontes que vertem, junto às quais a água através dos tubos, ou canais é conduzida; devendo desta forma, conforme a Teoria do movimento dos fluidos, a água inteiramente para a mesma altitude afluir, a partir da qual escoando em direção ao orifício tinha caído da fonte, entretanto a experiência atesta que esta altitude ela nunca atinge, mas a partir desta ela costuma a afastar-se mais, quanto mais longo o percurso pelos tubos tenha completado, e quanto mais estreitos tenham sido aqueles tubos.

[4] Nota do Tradutor: TABELA

2. Embora por causa disto a Teoria por muitas práticas não costuma parecer um tanto suspeita, ainda que ela seja suportada com os certíssimos princípios da mecânica, a sua verdade por causa deste dissenso é muito pouco debilitada, mas antes a causa desta aberração para estes tipos de circunstâncias convém ser procurada, as quais na Teoria não têm sido devidamente consideradas. Se porventura em realidade a água em seu movimento encontre impedimentos deste tipo, dos quais na Teoria nenhum cálculo tenha sido considerado, não é surpresa, se o resultado da Teoria muito pouco esclareça.

3. Mas se, por outro lado, esta diminuição de movimento, que a água sofre nos tubos, nós examinamos cuidadosamente, sem que nenhuma dúvida seja deixada para trás, porque então não seja esta iniciada pela fricção, ou pelo atrito da água contra a parede do tubo. Quando efetivamente corpos sólidos, à medida que são movidos sobre planos ainda que polidos, encontrem considerável resistência devido à fricção, efeito similar no movimento dos fluidos, enquanto nas paredes dos tubos, através dos quais transitam, são atritados, deve originar-se (a fricção), de onde o movimento do fluido seja retardado: de onde também é imediatamente percebido, este retardamento deve ser tanto maior, quanto maior a extensão que a água seja obrigada a percorrer nos tubos, e ao mesmo tempo quão estreitos tenham sido esses tubos.

4. Ainda que, no entanto, ninguém a tenha facilmente negado, porque então o movimento dos fluidos não seja da mesma maneira submetido à fricção, e até mesmo a maior parte dos Autores que anteriormente deram atenção ao mencionado enfraquecimento do movimento, atribuam o mesmo pela evidente fricção, entretanto nenhum deles, tanto quanto certamente consta para mim, ou determinou as leis desta fricção, ou pelo menos tenha tentado investigar. Tanto, por causa disto, seja esta determinação de grande importância, caso queiramos aplicar a Teoria na prática, quanto por causa daquelas grandes dificuldades que esteja envolvido oferecerei esta obra, de tal maneira que a matéria a tratar possa ser feita, por conta das forças que eu descubra e também que eu esclareça.

5. E primeiramente na verdade, não existindo nenhuma dúvida, de que as menores partículas dos fluidos possam ser retidas sobre sólidos, elas, enquanto avançam próximo a lateral dos tubos, e por aí quase comprimem, devem sentir o efeito da fricção; e também as leis desta fricção exatamente similares existirão para elas, as quais no movimento dos corpos sólidos são observadas, ainda que se a magnitude da fricção, por causa da maior lubricidade das partículas fluidas, sem dúvida seria muito menor.

6. Nos corpos principalmente nos sólidos observamos, a fricção, que se distribui por qualquer caminho sobre uma superfície, sempre dada a segurar em proporção à força, que por aí comprimem a superfície, de tal forma que nem a forma dos corpos, nem a extensão da base, que toquem a superfície, nenhum que confira aumento ou diminuição da fricção. Deste modo com experimentos foi verificado, se quaisquer corpos caírem sobre superfícies lenhosas, ou metálicas, contanto que não muito notável aspereza esteja presente, a fricção da quarta e sucessivamente da terça parte da força, as quais contra estas superfícies comprimem, eram as mesmas.

7. É, portanto, visto como certo, que é válida similar lei da fricção nos fluidos, de tal forma que para qualquer porção do fluido a fricção tenha certa proporção, diante da força, que esta porção do fluido contra as laterais do tubo, através do qual escoa, é comprimida: e aqui o cálculo por meio dos experimentos deverá ser explorado, o qual talvez possa ter sucesso em variações diversas, como tubos diferentes e também com matéria diferente que tenham sido produzidos. No cálculo, por conseguinte, será conveniente assumir que esta (a matéria) é indefinida, a fim de que ela mais tarde através dos experimentos, para os quais a Teoria será aplicada, possa ser determinada.

8. Se então nós colocamos (Fig. 1) a massa da água para escoar através do tubo $ABCD$ e as paredes internas do tubo sobre o elemento $MNmn$ pela água com certa força é premido, tanto mais fossem (as paredes) pressionadas, se sob a água quiescente imersa estivessem na profundidade $= p$, esta profundidade explicará o estado de compressão da água que estará aplicado contra o elemento $MNmn$. Mas se então $u$ exprime o perímetro da secção do tubo feita em $MN$ e $ds$ a extensão $Mm$ do elemento $MNmn$, será a superfície interna deste elemento $= uds$ e, portanto, a pressão, a qual aquela superfície suporta, será equacionada para o peso do volume de água $= puds$.

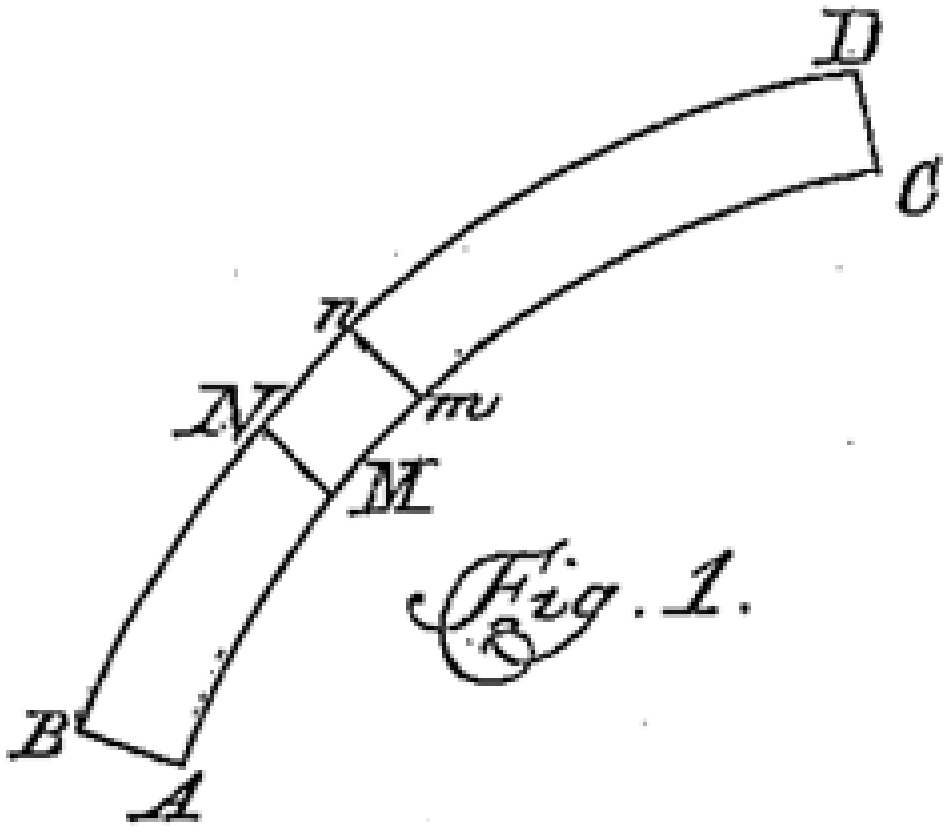

Fig. 1.

9. Já a fricção, a qual o elemento de água $MNmn$ sofre no seu movimento através do tubo, tendo uma dada proporção em relação à força de compressão, será indicada a força de fricção por meio de $\lambda puds$, onde $\lambda$ pode ser facilmente coletável na fricção de pouca importância, cujo valor deve ser determinado através dos experimentos. Desta forma, portanto, com força $= \lambda puds$ será contido o movimento do elemento de água $MNmn$, e seja a massa do elemento, dada pela amplitude do tubo dentro de $MN = zz$, $= zzds$, será o tal retardamento originado pela fricção

$$= \frac{\lambda pu}{zz}.$$

10. Consequentemente ponderará a fricção tanto o tamanho como a forma da cavidade do tubo: se para a secção do tubo em $MN$ fixemos que fosse feita retangular, com um lado sendo $= m$, outro $= n$: será o perímetro dela $u = 2m + 2n$, e com área $zz = mn$, de onde para este caso será o retardamento

$$= \frac{2\lambda p(m+n)}{mn},$$

e se a secção seja quadrada, ou $m = n$, será o retardamento

$$= \frac{4\lambda p}{n}.$$

Mas se, no entanto, seja circular, possuindo raio $=\ n$, por isso

$$u = 2\pi n \qquad \text{e} \qquad zz = \pi nn$$

será o retardamento

$$= \frac{2\lambda p}{n}.$$

11. E se por causa disso, posto que principalmente como é de costume, a secção do tubo seja circular e de sua área em $MN$ seja colocada $= zz$, será o raio dela $n = z/\sqrt{\pi}$, de onde o retardamento, originado pela fricção, revela

$$= \frac{2\lambda p\sqrt{\pi}}{z}.$$

Sendo, no entanto, $\pi = 3{,}14159265$ este número poderá ser combinado conjuntamente dentro de $2\lambda$, então a fim de que o retardamento originado pela fricção possamos exprimir por $\alpha p/z$: que em presença de experimentos, os quais na verdade de tal maneira são instituídos aos tubos, tornará o valor do próprio $\alpha$ conhecido, e sem muito recupera daí o próprio valor de $\lambda = \alpha/2\sqrt{\pi}$.

12. Portanto em relação ao efeito da fricção para onde quer que seja necessário investigar delimitando a queda da pressão, a qual a água apresenta em todos os lugares nas laterais: ou o estado de compressão da água em todos os locais do tubo se deve definir. Mas estando pendente o estado de compressão da velocidade, a velocidade na realidade seja diminuída pela fricção e por esta razão não seja possível sem ser ela (a velocidade) conhecida, é evidente esta investigação, segundo os bons costumes junto aos Analistas, é dever ser estabelecido, sem dúvida a fim de que as grandezas desconhecidas no cálculo bem como as conhecidas sejam tratadas.

13. Portanto eu remontarei esta investigação desde os primeiros princípios da mecânica com a qual se possa compreender facilmente a verdade das conclusões a partir desse momento deduzidas. Seja então (Fig. 2) um vaso no alto em $AB$ constantemente cheio com água, já seja porque a amplitude $AB$ seja infinita, já seja porque continuamente um adequado recurso d´água flua. Além disso, este vaso acaba na parte de baixo num tubo, ou melhor, no canal $ABMNCD$, tanto de forma quanto de amplitude variável de qualquer modo, através do qual a água escoe e em seu orifício $CD$ irrompa.

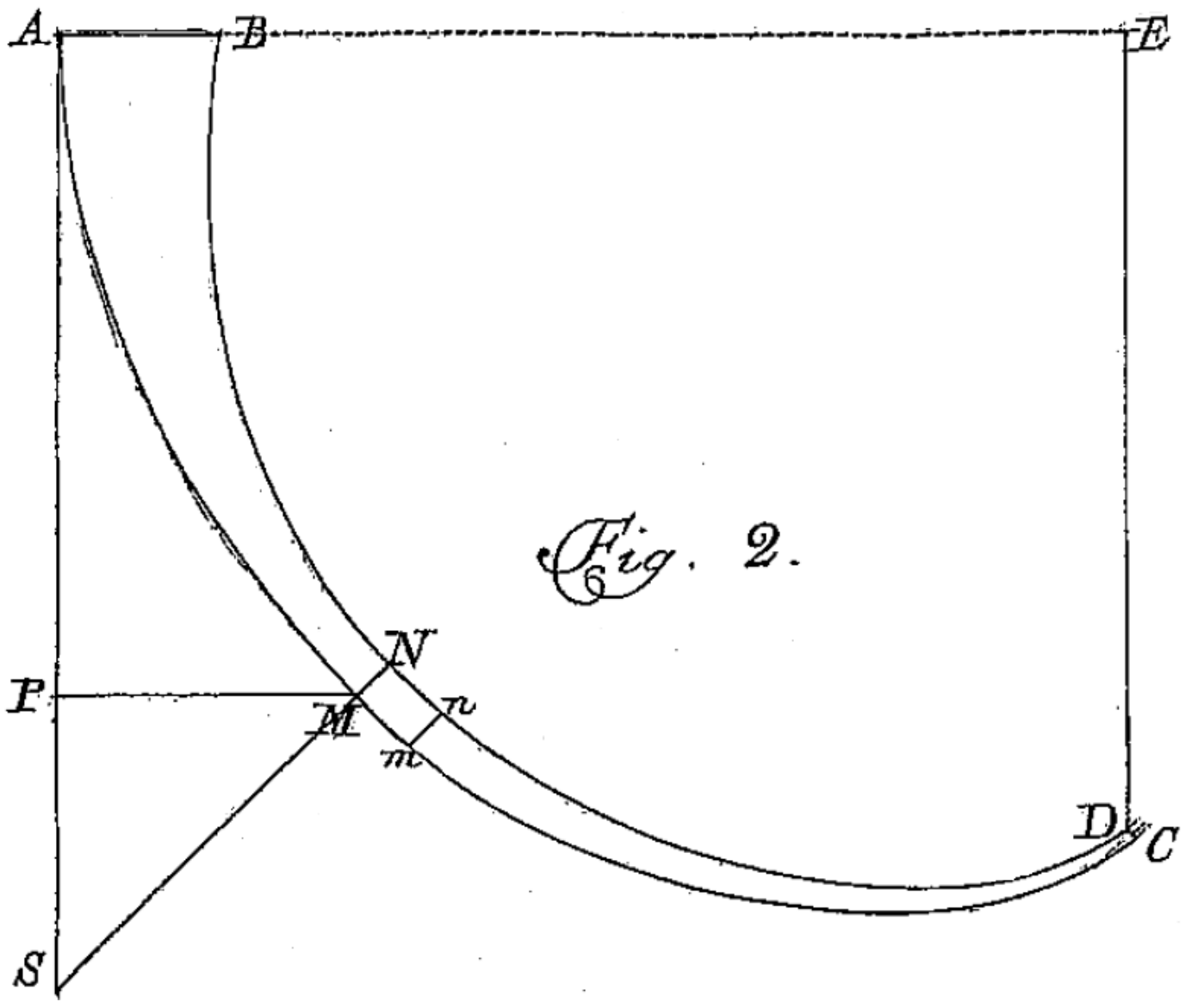

14. Mesmo se o movimento da água, tendo começado a fluir pela primeira vez, é acelerada, ainda que em pouco tempo for reconduzido para um movimento uniforme, com o qual logo em seguida prosseguirá a fluir continuamente. Por causa disto, nesta coisa, assumirei que o movimento da água neste momento chegou até este estado de uniformidade, de tal sorte que a questão seria com quanta velocidade a água teria sido expelida do orifício $CD$; ou com a fricção removida, a velocidade da água é para a altitude $DE$, que é rebaixada como se um orifício estivesse embaixo da superfície superior da água $AB$, pede-se, quanto menor seja então a velocidade futura por causa da fricção.

15. Seja então a amplitude do orifício $CD = hh$, e a velocidade, por onde esta água no ar é irrompida, seja devida à altitude $= v$, será, portanto, $v$ uma quantidade fixa. Seja ainda a profundidade deste orifício $CD$ sob o nível da água $ABE$, ou altitude $DE = a$. Então sendo guiada a reta para a vertical $APS$, um ponto qualquer $M$ do tubo seja referenciado às coordenadas ortogonais $AP = x$ e $PM = y$, em $M$ na verdade seja a amplitude do tubo $MN = zz$, e será a velocidade da água por esta secção $MN$ limitada por

$$= \frac{hh\sqrt{v}}{zz},$$

ou devida à altitude

$$= \frac{h^4 v}{z^4};$$

pela razão que as velocidades da água são recíprocas para as amplitudes[5] do tubo.

16. Coloquemos um diferencial do tempo $dt$ suficiente para que uma gota d´água aparecida em $MN$ chegue à secção $mn$ e seja o elemento

$$Mm = \surd(dx^2 + dy^2) = ds,$$

o qual um pequeno espaço percorrido no diferencial de tempo $dt$ pela velocidade $(hh\surd v)/zz$, será

$$ds = \frac{hhdt\surd v}{zz} \text{ou} dt = \frac{zzds}{hh\surd v} \ ,$$

onde deve ser notado que irão ser s e $zz$ funções das próprias coordenadas $x$ e $y$, com as quais é designado o ponto $M$ do tubo. Além do mais coloquemos o elemento $Mm$ inclinado com o ângulo $= \varphi$ em direção à reta vertical, será

$$dx = ds \cos\varphi \qquad \text{e} \qquad dy = ds \cos\varphi.$$

17. Contudo é preciso que a água tenha a força de manter este movimento que concebemos, posto que qualquer gota d'água, contida na secção $MN$, que seja impelida por dupla força aceleradora, das quais uma é vertical segundo $AP$

$$= \frac{2ddx}{dt^2},$$

a outra certamente horizontal segundo $PM$

$$= \frac{2ddy}{dt^2},$$

aplicado com o elemento $dt$ fixo. Todavia é

[5] Edição primeira: altitudes.

$$\frac{dx}{dt} = \frac{hh \cos\varphi \ \sqrt{v}}{zz} \text{ e } \quad \frac{dy}{dt} = \frac{hh\, sen\, \varphi \sqrt{v}}{zz};$$

donde é feito:

$$\frac{ddx}{dt} = -\frac{hh\, d\varphi sen\, \varphi \sqrt{v}}{zz} - \frac{2hh\, dz \cos\varphi \ \sqrt{v}}{z^3}$$

e

$$\frac{ddy}{dt} = -\frac{hhd\varphi \cos\varphi \ \sqrt{v}}{zz} - \frac{2hh\, dz\, sen\, \varphi \sqrt{v}}{z^3}.$$

18. Estas fórmulas multiplicadas por

$$\frac{2}{dt} = \frac{2hh\sqrt{v}}{zzds}$$

fornecerão as forças aceleradoras procuradas

$$\text{Força } \ AP = h^4 v \left(-\frac{2\, d\varphi sen\, \varphi}{z^4 ds} - \frac{4\, dz \cos\varphi}{z^5 ds}\right),$$

$$\text{Força } \ PM = h^4 v \left(\frac{2d\varphi \cos\varphi}{z^4 ds} - \frac{4\, dz\, sen\, \varphi}{z^5 ds}\right).$$

Donde além do mais são derivadas duas outras forças segundo as direções $Mm$ e $MS$, das quais esta é normal àquela. Surge, entretanto

$$\text{Força } \ Mm = \text{força } AP \cos\varphi + \text{força } PM\, sen\, \varphi \ = -\frac{4\, h^4 v\, dz}{z^5 ds},$$

$$\text{Força } \ MS = \ \text{força } AP\, sen\, \varphi - \text{força } PM \cos\varphi = -\frac{2\, h^4 v\, d\varphi}{z^4 ds}.$$

19. No presente estágio, temos apenas uma dificuldade para a força anteriormente exposta, com a qual a água é propelida segundo a direção $Mm$: e, portanto, esta força aceleradora deve ser igual à mesma força, com a qual a água é efetivamente posta em movimento no tubo segundo esta direção. Mas primeiro, em qualquer lugar, uma gotícula de água é forçada para baixo pela força da gravidade, que ao manifestar-se com uniformidade, dará origem, portanto, a uma força aceleradora segundo a direção do tubo $Mm = cos\varphi$.

20. Então se o estado de compressão da água em $MN$ for expresso pela altitude $p$, será ele em $mn = p + dp$: logo sendo a água do elemento $MNmn$ propelida para frente pela força $p$, mas sendo repelida pela força $p + dp$ para trás, logo originará uma força aceleradora dirigida segundo a direção $Mm$

$$= -\frac{dp}{ds},$$

sendo $p$ função do próprio $s$ ou das mesmas $x$ e $y$.

21. Terceiro por causa da fricção o movimento da água é resistido pela força retardadora $= \frac{(\propto p)}{z}$, como foi desenvolvida acima no parágrafo 11, donde a força aceleradora segundo a direção $Mm$ será

$$= -\frac{\propto p}{z}.$$

Com o resultado da água estando submetida a estas três forças, é necessário que seja:

$$-\frac{4\,h^4 v\,dz}{z^5 ds} = \cos\varphi - \frac{dp}{ds} - \frac{\propto p}{z}$$

ou

$$dx - dp - \frac{\propto p\,ds}{z} + \frac{4\,h^4 v\,dz}{z^5} = 0,$$

por causa de $ds\ cos\varphi = dx$.

22. Chegamos, portanto, a esta equação, de onde em primeiro lugar deve ser definido o estado de compressão para qualquer região $MN$ do tubo

$$dp + \frac{\propto p\,ds}{z} = dx + \frac{4\,h^4 v\,dz}{z^5},$$

que multiplicada por $e^{\propto\int\frac{ds}{z}}$, denotando $e$ o número, cujo logaritmo hiperbólico é $= 1$, se torna integrável. Por conta da brevidade será considerado como estabelecido $\int\frac{ds}{z} = r$:

$$e^{\propto r}p = C + \int e^{\propto r}dx + 4\, h^4 v \int \frac{e^{\propto r}dz}{z^5}\,.$$

23. Mas o valor desta integral $\int \frac{ds}{z} = r$ seja deste modo aceito, para que na região mais elevada $AB$ do vaso, ou onde $x = 0$, se anule. Com isto posto, sendo $\propto$ uma fração extremamente pequena, será

$$e^{\propto r} = 1{+}\propto r + \frac{1}{2}\propto^2 r^2 + \frac{1}{6}\propto^3 r^3 + \frac{1}{24}\propto^4 r^4 + \text{etc. },$$

será

$$e^{\propto r}p = C + x{+}\propto \int rdx + \frac{1}{2}\propto^2 \int rrdx + \frac{1}{6}\propto^3 \int r^3dx$$

$$-\frac{h^4v}{z^4} - 4 \propto h^4v \int \frac{rdz}{z^5} - 2 \propto^2 h^4v \int \frac{rrdz}{z^5} - \frac{2}{3}\propto^3 h^4v \int \frac{r^3dz}{z^5} - \text{ etc.}$$

24. Sendo $dr = \frac{ds}{z}$, com isso a integração também é possível de ser realizada com método instituído:

$$\int e^{\propto r}dx = e^{\propto r}x - \propto \int \frac{e^{\propto r}xds}{z}\,,$$

$$\int \frac{e^{\propto r}dz}{z^5} = -\frac{e^{\propto r}}{4z^5} + \frac{\propto}{4}\int \frac{e^{\propto r}ds}{z^5}$$

e assim desta forma será

$$p = Ce^{-\propto r} + x{-}\propto e^{-\propto r}\int \frac{e^{\propto r}xds}{z} - \frac{h^4v}{z^4} + \propto h^4ve^{-\propto r}\int \frac{e^{\propto r}ds}{z^5}$$

ou

$$p = +C{-}\propto Cr + \frac{1}{2}\propto\propto Crr - \frac{1}{6}\propto^3 Cr^3$$

$$+x{-}\propto \int \frac{xds}{z} {+}\propto\propto r \int \frac{xds}{z} - \frac{1}{2}\propto^3 rr \int \frac{xds}{z}$$

$$-\frac{h^4v}{z^4} -^{6} \propto h^4v \int \frac{ds}{z^5} {-}\propto\propto \int \frac{rxds}{z} + \propto^3 r \int \frac{rxds}{z}$$

[6] Edição primeira: $+$. Tradutor Corrigiu

$$+\propto\propto h^4vr\int\frac{ds}{z^5}-\frac{1}{2}\propto^3\int\frac{rrxds}{z}$$

$$-^{7}\propto\propto h^4v\int\frac{rds}{z^5}-{}^{8}\frac{1}{2}\propto^3 h^4vrr\int\frac{ds}{z^5}$$

$$+^{9}\propto^3 h^4vr\int\frac{rds}{z^5}$$

$$-^{10}\frac{1}{2}\propto^3 h^4v\int\frac{rrds}{z^5}+\text{etc.}$$

25. Também explicará a partir de onde desenvolvi alguns casos particulares, através dos quais o efeito da fricção se pode mostrar facilmente. Seja, portanto (Fig. 3), o vaso cilíndrico vertical mais elevado $ABEF$, cuja amplitude seja $= gg$ e a altura $AE = a$; além disso, acoplado a este vaso, exista o tubo cilíndrico $FCD$, de comprimento $DF = b$ e de amplitude $= ff$, que faça com a reta vertical o ângulo $= \xi$; também na base inferior exista com abertura $CD = hh$, através da qual a água escoe para fora, enquanto o vaso superior é mantido continuamente cheio.

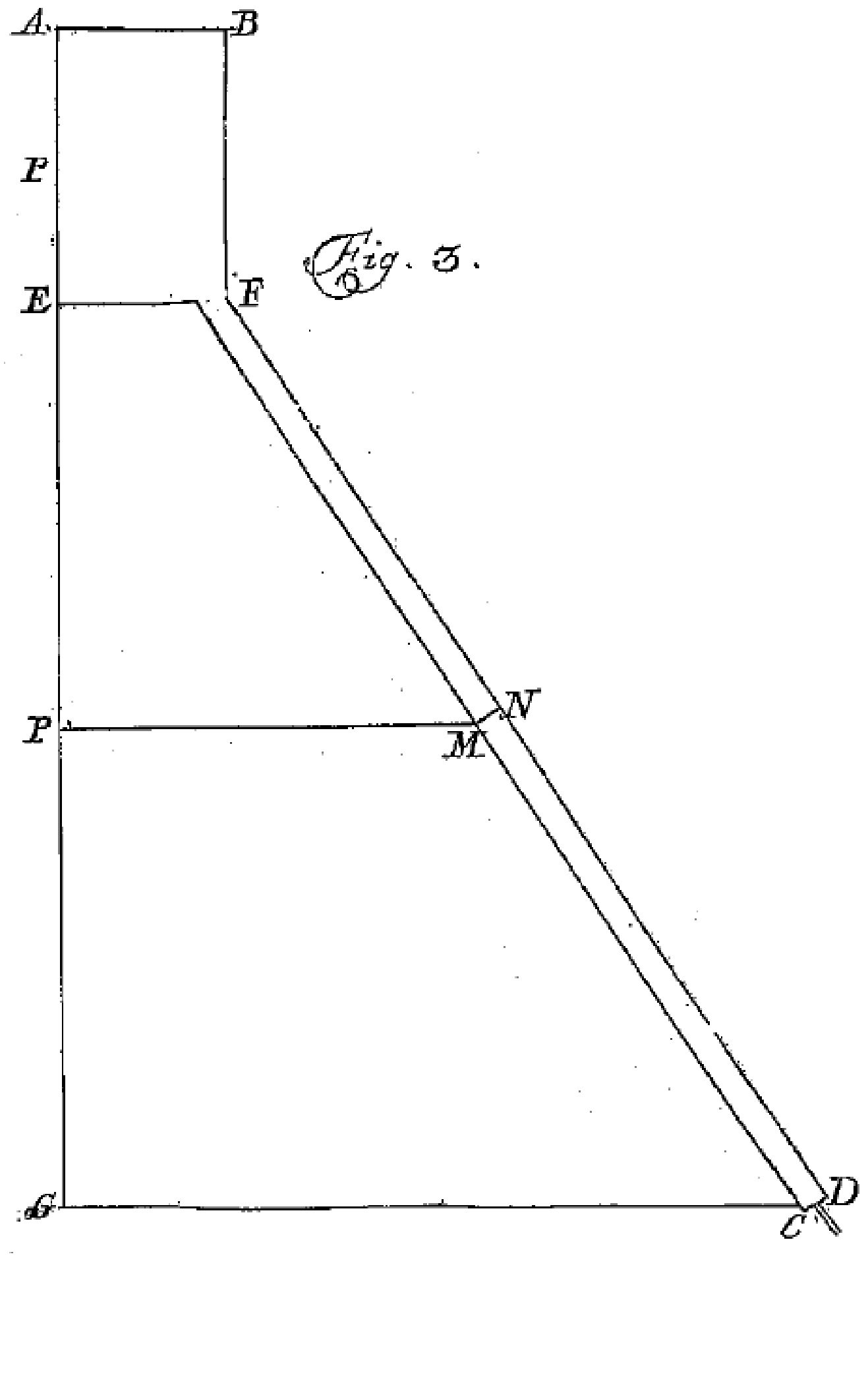

26. Tomemos primeiro o ponto indefinido $P$ no tubo

[7] Edição primeira: $+$. Tradutor Corrigiu

[8] Edição primeira: $+$. Tradutor Corrigiu

[9] Edição primeira: $-$. Tradutor Corrigiu

[10] Edição primeira: $+$. Tradutor Corrigiu

superior $ABEF$, onde é $z = g$, $ds = dx$ e $r = \frac{x}{g}$. Logo de tal modo integrando, para que a integral no ponto supremo $A$ desapareça, será:

$$\int e^{\propto r} dx = \int e^{\frac{\propto x}{g}} dx = \frac{g}{\alpha}\left(e^{\frac{\propto x}{g}} - 1\right),$$

$$4\int \frac{e^{\propto r} dz}{z^5} = \frac{1 - e^{\propto r}}{g^{4^{11}}} + \frac{\alpha}{g^5}\int e^{\frac{\propto x}{g}} dx$$

e por esta razão

$$4\int \frac{e^{\propto r} dz}{z^5} = -\frac{1}{g^4}\left(e^{\frac{\propto x}{g}} - 1\right) + \frac{1}{g^4}\left(e^{\frac{\propto x}{g}} - 1\right) = 0\,,$$

portanto

$$e^{\frac{\alpha x}{g}} p = C + \frac{g}{\alpha}\left(e^{\frac{\propto x}{g}} - 1\right)$$

ou

$$p = Ce^{\frac{-\alpha x}{g}} + \frac{g}{\alpha}\left(1 - e^{\frac{-\propto x}{g}}\right).$$

27. Sendo também em geral

$$4\int \frac{e^{\propto r} dz}{z^5} = \frac{1}{g^4} - \frac{e^{\propto r}}{z^4} + \propto \int \frac{e^{\propto r} ds}{z^5},$$

será no ponto extremo $E$ dos vasos, onde a amplitude seja subitamente $= ff$

$$4\int \frac{e^{\propto r} dz}{z^5} = \frac{1}{g^4} - \frac{1}{f^4} e^{\frac{\propto x}{g}} + \frac{1}{g^4}\left(e^{\frac{\propto x}{g}} - 1\right) = e^{\frac{\propto x}{g}}\left(\frac{1}{g^4} - \frac{1}{f^4}\right),$$

de onde seja a pressão em $F$

$$= Ce^{\frac{-\alpha a}{g}} + \frac{g}{\alpha}\left(1 - e^{\frac{-\propto a}{g}}\right) + h^4 v\left(\frac{1}{g^4} - \frac{1}{f^4}\right).$$

---

[11] Edição primeira: $g^5$. Tradutor Corrigiu

Mas na superfície suprema $AB$ a pressão será $=\ C$, que deverá ser igual à pressão da atmosfera, é a mesma equacionada à coluna de água de altitude $=\ l$, será $C = l$ e a pressão em $F$

$$= le^{\frac{-\alpha a}{g}} + \frac{g}{\alpha}\left(1 - e^{\frac{-\propto a}{g}}\right) + h^4 v\left(\frac{1}{g^4} - \frac{1}{f^4}\right).$$

28. Com a pressão em $F$ descoberta, consideremos agora exclusivamente outro tubo $FCD$, para o qual é

$$zz = ff \qquad \text{e} \qquad dx = ds\, cos\zeta\ .$$

Seja também agora $EP = x$ e a pressão em $M = p$. Agora por causa de $r = \frac{s}{f}$ será

$$\int e^{\propto r} dx = \frac{f}{\alpha}\left(e^{\frac{\propto s}{f}} - 1\right) cos\zeta$$

e

$$4\int \frac{e^{\propto r} dz}{z^5} = \frac{1}{f^4} - \frac{e^{\frac{\propto s}{f}}}{z^4} + \frac{1}{f^4}\left(e^{\frac{\propto s}{f}} - 1\right),$$

cujo valor desaparece em qualquer ponto intermediário, mas no ponto $D$ no entanto, onde subitamente $z = h$ e $s = b$, será

$$4\int \frac{e^{\propto r} dz}{z^5} = e^{\frac{\propto b}{f}}\left(\frac{1}{f^4} - \frac{1}{h^4}\right).$$

29. Portanto em qualquer ponto intermediário $M$ será a pressão

$$p = Ce^{\frac{-\alpha s}{f}} + \frac{f}{\alpha}\left(1 - e^{\frac{-\propto s}{f}}\right) cos\zeta\ ,$$

e porque, tendo sido colocado $s = 0$, a pressão em $F$ revela-se $= C$, é necessário, que seja

$$C = le^{\frac{-\alpha a}{g}} + \frac{g}{\alpha}\left(1 - e^{\frac{-\propto a}{g}}\right) + h^4 v\left(\frac{1}{g^4} - \frac{1}{f^4}\right).$$

Com este valor escrito a pressão se revela no orifício extremo $CD$

$$= Ce^{\frac{-\alpha b}{f}} + \frac{f}{\alpha}\left(1 - e^{\frac{-\propto b}{f}}\right) cos\zeta + h^4 v\left(\frac{1}{f^4} - \frac{1}{h^4}\right).$$

30. Visto que esta água verdadeiramente irrompe na atmosfera, outra pressão não suporta, além do peso da atmosfera, donde será

$$l = Ce^{\frac{-\alpha b}{f}} + \frac{f}{\alpha}\left(1 - e^{\frac{-\propto b}{f}}\right) \cos\zeta + h^4 v\left(\frac{1}{f^4} - \frac{1}{h^4}\right)$$

e substituído pelo valor do próprio $C$

$$l = le^{\frac{-\alpha a}{g} - \frac{\alpha b}{f}} + \frac{g}{\alpha} e^{-\frac{\alpha b}{f}}\left(1 - e^{\frac{-\propto a}{g}}\right) + e^{-\frac{\alpha b}{f}} h^4 v\left(\frac{1}{g^4} - \frac{1}{f^4}\right)$$
$$+ \frac{f}{\alpha}\left(1 - e^{\frac{-\propto b}{f}}\right) \cos\zeta + h^4 v\left(\frac{1}{f^4} - \frac{1}{h^4}\right).$$

31. Imediatamente desta equação é extraída a velocidade, pela qual a água irrompe através do orifício $CD$. Assim aparecerá a altura devida por esta velocidade

$$v = \frac{l\left(1 - e^{\frac{-\alpha a}{g} - \frac{\alpha b}{f}}\right) - \frac{g}{\alpha} e^{-\frac{\alpha b}{f}}\left(1 - e^{\frac{-\propto a}{g}}\right) - \frac{f}{\alpha}\left(1 - e^{\frac{-\propto b}{f}}\right)\cos\zeta}{e^{-\frac{\alpha b}{f}} \cdot \frac{h^4}{g^4} + \left(1 - e^{\frac{-\propto b}{f}}\right)\frac{h^4}{f^4} - 1}$$

ou sendo mudados os sinais do numerador e do denominador

$$v = \frac{\frac{g}{\alpha} e^{-\frac{\alpha b}{f}}\left(1 - e^{\frac{-\propto a}{g}}\right) + \frac{f}{\alpha}\left(1 - e^{\frac{-\propto b}{f}}\right)\cos\zeta - l\left(1 - e^{\frac{-\alpha a}{g} - \frac{\alpha b}{f}}\right)}{1 - \left(1 - e^{\frac{-\propto b}{f}}\right)\frac{h^4}{f^4} - e^{-\frac{\alpha b}{f}} \cdot \frac{h^4}{g^4}}.$$

Mas se consequentemente o tubo $FD$ seja vertical, será feito $cos\zeta = 1$; mas se, no entanto, este fosse horizontal, será $cos\zeta = 0$.

32. Por causa de $\alpha$ sendo extremamente pequeno,

$$e^{\frac{-\alpha a}{g}} = 1 - \frac{\propto a}{g} + \frac{\propto^2 a^2}{2g^2} - \frac{\propto^3 a^3}{6g^3} + etc.\ ,$$

$$e^{\frac{-\alpha b}{f}} = 1 - \frac{\propto b}{f} + \frac{\propto^2 b^2}{2f^2} - \frac{\propto^3 b^3}{6f^3} + etc.\ ,$$

será, não avançando além da segunda potência do próprio $\alpha$:

$$v = \frac{\begin{cases} a + b\cos\zeta \\ -\propto\left(\frac{aa}{2g} + \frac{ab}{f} + \frac{bb\cos\zeta}{2f} + \frac{al}{g} + \frac{bl}{f}\right) \\ +\propto\propto\left(\frac{a^3}{6gg} + \frac{aab}{2fg} + \frac{abb}{2ff} + \frac{b^3\cos\zeta}{6ff} + \frac{1}{2}l\left(\frac{a}{g} + \frac{b}{f}\right)^2\right) \end{cases}}{1 - \frac{h^4}{g^4} - \frac{\alpha b}{f}\left(\frac{h^4}{f^4} - \frac{h^4}{g^4}\right) + \frac{\alpha\alpha bb}{2ff}\left(\frac{h^4}{f^4} - \frac{h^4}{g^4}\right)}$$

onde $a + b\ cos\xi$ denota a altitude total $AG$.

33. Se a amplitude do vaso superior $ABEF$ for quase infinita ou $g = \infty$, será:

$$v = \frac{a + b\ cos\zeta - \frac{\alpha b}{f}\left(a + \frac{1}{2}b\cos\zeta + l\right) + \frac{\alpha^2 bb}{2ff}\left(a + \frac{1}{3}b\cos\zeta\ l\right)}{1 - \frac{\alpha b h^4}{f^5} + \frac{\alpha\alpha bbh^4}{2f^6}}$$

donde fica exposto que a velocidade é assim menor, que aquela que adquire o corpo caindo da altitude $AG$; e também especialmente uma diminuição que aplica o peso $l$ da atmosfera, deste modo ainda que no vácuo o efeito da fricção há-de ser muito menor.

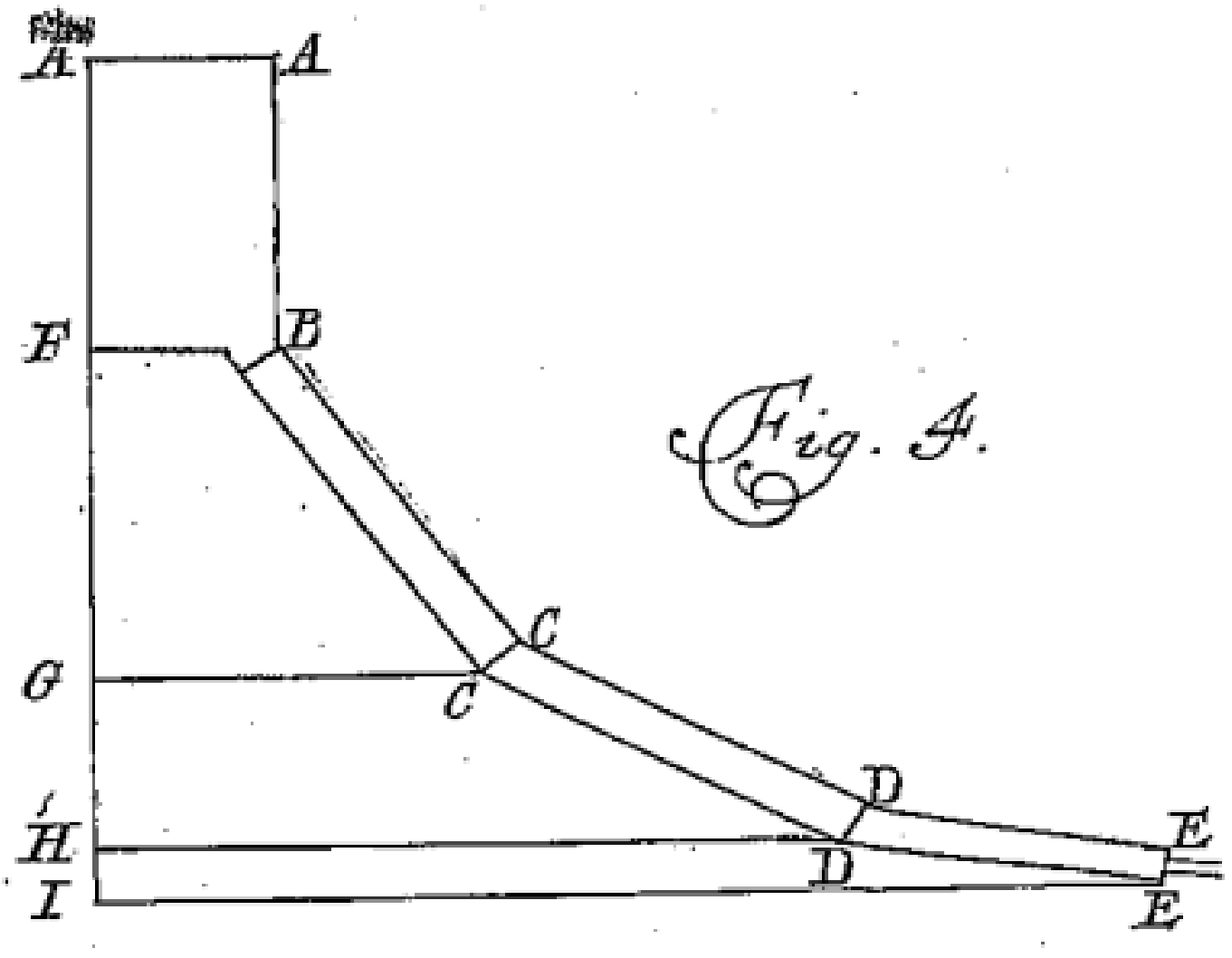

34. Se o canal (Fig. 4) através do qual a água cai, se ajusta por meio de vários tubos cilíndricos inclinados de qualquer modo em relação ao horizonte, então o movimento da água ou a velocidade do efluxo, com o movimento tendo sido prontamente restaurado à uniformidade, sem dificuldade poderá ser resolvido.

Seja, pois, estabelecido para cada uma das partes

$AF = a$, amplitude $AA = ff$ e ângulo com a vertical $= 0$ ,
$BC = b$, amplitude $BB = gg$ e ângulo com a vertical $= \zeta$ ,
$CD = c$, amplitude $AA = hh$ e ângulo com a vertical $= \eta$ ,
$DE = d$, amplitude $DD = ii$ e ângulo com a vertical $= \theta$ .

Finalmente seja o orifício $EE = kk$, que até aqui indicamos por $hh$.

35. Seja enfim $v$ a altitude devida pela velocidade, com que a água efluirá através do orifício $EE = kk$, e seja colocado o estado de compressão da água em $AA = l$ , que é uma altitude em torno de 30 pés,
em $BB = P$ ,
em $CC = Q$ ,
em $DD = R$ ,
em $EE = l$ , com este seja feito o efluxo.

36. Mas posto que o cálculo já estabelecemos antes, encontraremos

$$P = le^{\frac{-\alpha a}{f}} + \frac{f}{\alpha}\left(1 - e^{\frac{-\propto a}{f}}\right) \qquad + k^4 v\left(\frac{1}{f^4} - \frac{1}{g^4}\right),$$

$$Q = Pe^{\frac{-\alpha b}{g}} + \frac{g}{\alpha}\left(1 - e^{\frac{-\propto b}{g}}\right) cos\,\zeta + k^4 v\left(\frac{1}{g^4} - \frac{1}{h^4}\right),$$

$$R = Qe^{\frac{-\alpha c}{h}} + \frac{h}{\alpha}\left(1 - e^{\frac{-\propto c}{h}}\right) \cos\eta + k^4 v\left(\frac{1}{h^4} - \frac{1}{i^4}\right),$$

$$l = Re^{\frac{-\alpha d}{i}} + \frac{i}{\alpha}\left(1 - e^{\frac{-\propto d}{i}}\right) \cos\theta + k^4 v\left(\frac{1}{i^4} - \frac{1}{k^4}\right)$$

e daqui a velocidade procurada ou a altitude $v$ é definida: e simultaneamente estabelece a lei, que servirá para avançar, se o canal é composto com várias partes.

37. A fim de que desenvolvamos estas fórmulas com comodidade, estabeleçamos em favor da brevidade

$$e^{\frac{-\propto a}{f}} = 1 - \propto A, \qquad e^{\frac{-\propto b}{g}} = 1 - \propto B,$$

$$e^{\frac{-\propto c}{h}} = 1 - \propto C, \quad e^{\frac{-\propto d}{i}} = 1 - \propto D,$$

de tal modo que seja

$$A = \frac{a}{f}\left(1 - \frac{\alpha a}{2f} + \frac{\alpha^2 a^2}{6f^2} - \frac{\alpha^3 a^3}{24f^3} + etc.\right),$$

$$B = \frac{b}{g}\left(1 - \frac{\alpha b}{2g} + \frac{\alpha^2 b^2}{6g^2} - \frac{\alpha^3 b^3}{24g^3} + etc.\right),$$

$$C = \frac{c}{h}\left(1 - \frac{\alpha c}{2h} + \frac{\alpha^2 c^2}{6h^2} - \frac{\alpha^3 c^3}{24h^3} + etc.\right),$$

$$D = \frac{d}{i}\left(1 - \frac{\alpha d}{2i} + \frac{\alpha^2 d^2}{6i^2} - \frac{\alpha^3 d^3}{24i^3} + etc.\right).$$

38. Sendo consequentemente

$$P = (1 - \propto A)l + Af + k^4 v\left(\frac{1}{f^4} - \frac{1}{g^4}\right),$$

será

$$Q = (1 - \propto A)(1 - \propto B)l + (1 - \propto B)Af + (1 - \propto B)k^4 v\left(\frac{1}{f^4} - \frac{1}{g^4}\right)$$
$$+ Bg\cos\zeta + k^4 v\left(\frac{1}{f^4} - \frac{1}{g^4}\right)$$

ou

$$Q = (1 - \propto A)(1 - \propto B)l + (1 - \propto B)Af + Bg\cos\zeta$$
$$+(1 - \propto B)\frac{k^4 v}{f^4} + \propto B\frac{k^4 v}{f^4} - \frac{k^4 v}{h^4}.$$

Daqui em diante é feito

$$R = (1 - \propto A)(1 - \propto B)(1 - \propto C)l + A(1 - \propto B)(1 - \propto C)f$$
$$+B(1 - \propto C)g\cos\zeta + Ch\cos\eta + (1 - \propto B)(1 - \propto C)\frac{k^4 v}{f^4}$$

$$+\propto B(1-\propto C)\frac{k^4 v}{g^4}+\propto C\frac{k^4 v}{h^4}-\frac{k^4 v}{i^4},$$

donde finalmente é obtida esta igualdade

$$l-(1-\propto A)(1-\propto B)(1-\propto C)(1-\propto D)l$$

$$=A(1-\propto B)(1-\propto C)(1-\propto D)f+B(1-\propto C)(1-\propto D)gcos\,\zeta$$

$$+C(1-\propto D)h\cos\eta+Di\cos\theta+(1-\propto B)(1-\propto C)(1-\propto D)\frac{k^4 v}{f^4}$$

$$+\propto B(1-\propto C)(1-\propto D)\frac{k^4 v}{h^4}+\propto C(1-\propto D)\frac{k^4 v}{h^4}+\propto D\frac{k^4 v}{i^4}-v\,.$$

39. Seja, além disso, colocado para abreviação

$$e^{\frac{-\propto a}{f}}=1-\propto A=\mathfrak{A},\quad e^{\frac{-\propto b}{g}}=1-\propto B=\mathfrak{B},$$

$$e^{\frac{-\propto c}{h}}=1-\propto C=\mathfrak{C},\quad e^{\frac{-\propto d}{i}}=1-\propto D=\mathfrak{D},$$

além disso verdadeiramente:

$$\frac{k^4}{f^4}=\mathfrak{f}\ ,\quad \frac{k^4}{g^4}=\mathfrak{g}\ ,\quad \frac{k^4}{h^4}=\mathfrak{h}\quad \text{e}\quad \frac{k^4}{i^4}=\mathfrak{i}\ ,$$

será

$$v=\frac{A\mathfrak{B}\mathfrak{C}\mathfrak{D}f+B\mathfrak{C}\mathfrak{D}g\ cos\zeta+C\mathfrak{D}h\ \cos\eta+\ Di\cos\theta\ -l+\mathfrak{A}\mathfrak{B}\mathfrak{C}\mathfrak{D}l}{1-\mathfrak{B}\mathfrak{C}\mathfrak{D}\mathfrak{f}-\propto B\mathfrak{C}\mathfrak{D}\mathfrak{g}-\propto C\mathfrak{D}\mathfrak{h}-\propto D\mathfrak{i}},$$

onde $a+b\cos\zeta+c\cos\eta+d\cos\theta$ é a altitude da água mais elevada acima do orifício $EE$.

40. Caso negligenciemos todas as potências do próprio $\propto$ superiores a primeira, encontraremos:

$$v = \frac{\left\{\begin{aligned} & a + b\cos\zeta + c\cos\eta + d\cos\theta - \propto a\left(\frac{a}{2f} + \frac{b}{g} + \frac{c}{h} + \frac{d}{i}\right) \\ & -\propto l\left(\frac{a}{f} + \frac{b}{g} + \frac{c}{h} + \frac{d}{i}\right) \qquad -\propto b cos\zeta\left(\frac{b}{2g} + \frac{c}{h} + \frac{d}{i}\right) \\ & \qquad\qquad\qquad\qquad\qquad -\propto c\cos\eta\left(\frac{c}{2h} + \frac{d}{i}\right) \\ & \qquad\qquad\qquad\qquad\qquad -\propto d\cos\theta\left(\frac{d}{2i}\right) \end{aligned}\right.}{1 - \mathfrak{f} + \propto \mathfrak{f}\left(\frac{b}{g} + \frac{c}{h} + \frac{d}{i}\right) - \propto \mathfrak{g}\frac{b}{g} - \propto \mathfrak{h}\frac{c}{h} - \propto \mathfrak{i}\frac{d}{i}}$$

e, se fosse exatamente nula a fricção ou $\propto = 0$, surgisse:

$$v = \frac{a + b\cos\zeta + c\cos\eta + d\cos\theta}{1 - \mathfrak{f}},$$

donde criada com a altitude total $AI = q$ será

$$v = \frac{f^4}{f^4 - k^4} \cdot q\,.$$

41. Em trânsito observo o seguinte, caso a amplitude dos vasos de cima seja máxima com relação a amplitude do orifício, haverá de ser $v = q$, ou a velocidade do efluxo em $EE$ é devida à altitude $AI = q$, o que concorda com o princípio conhecido. Mas se, no entanto, a amplitude dos vasos de cima $ff$ não muito exceda a amplitude do orifício $kk$, então $v > q$, ou se a água com maior velocidade efluisse, com relação ao caso precedente, o que por não pouco será percebido um paradoxo. A verdadeira razão desta aceleração na nossa hipótese deve ser procurada, por um lado assumimos que o vaso supremo é mantido sempre cheio com água, onde no cálculo colocamos que a água escoante corre naquele vaso com a mesma velocidade, que a água que nele entra: e deste modo esta água inicia o escoamento sem o repouso; donde não é surpresa, se ela irrompa através do orifício com maior velocidade, que aquela que convém à altitude $AI = q$.

42. Além disso, a partir daqui é percebido, porque o valor do próprio $v$, no caso em que o orifício $kk$ é igual ao da amplitude da suprema $ff$, até que se fez infinito; daqui evidentemente chegou ao conhecimento este caso do movimento da água ser acelerado continuamente e nem que em algum tempo alcance o estado de uniformidade. Há quanto tempo, pois, quantidade igual de água acima flui, quanto embaixo irrompe, e certamente sempre tanta velocidade, quanto a água que entra, este movimento continuo da mesma maneira será acelerado, e que costuma ocorrer na queda dos mais pesados. Além disso, muito menos estado de uniformidade local pode existir, se fosse $kk > ff$, exceto, para o caso que a água é separada das laterais do tubo.

43. Se também consequentemente o vaso supremo é mantido continuamente cheio com água, a não ser se ao mesmo tempo a água seja derramada no vaso com tanta velocidade, quanto cai a superfície suprema, o cálculo a partir da Teoria não é capaz de ser conduzido a um resultado. Mas se consequentemente desejemos acomodar o cálculo aos experimentos, será necessário ser aceito o vaso supremo amplíssimo, para que $kk$ diante de $ff$ possa ser rejeitado com segurança; assim efetivamente a água que é derramada vagarosamente o movimento da água não é turbilhonado, sendo que a superfície suprema irá também cair muito lentamente.

44. No entanto, para que reconheçamos com mais clareza o efeito da fricção, convirá ser desenvolvido alguns casos mais simples, que possam ser comparados com experimentos, para que então se possa definir o valor da letra $\propto$. No entanto, com este valor uma vez definido, deixei para trás todos os casos, quanto mais tenham requerido multiplicidade de esforços, é fornecida abundância de fórmulas livradas de dificuldade e principalmente será determinada a diminuição da velocidade oriunda da fricção: todavia devido à razão antes atribuída à amplitude suprema $ff$ diante do orifício $kk$ estabelecerei enfaticamente maior, para que assim o valor da letra $\mathfrak{f}$ pelo valor nulo possa ser tomado. No entanto é manifestado, contanto que seja $f > 3k$, haverá de ser $\mathfrak{f} < \frac{1}{81}$, cujo valor poderá ser desprezado sem erro apreciável.

## CASO I

## CASO A ÁGUA DO VASO SUPREMO EFLUA ATRAVÉS DE UM TUBO CILÍNDRICO VERTICAL

45. Seja (Fig. 5) a amplitude do vaso supremo $AA = ff$ e sua altura $AB = a$, que estabeleço ser mantido sempre cheio com água. A este vaso seria fixado verticalmente o tubo cilíndrico $BBCC$ de altura $BC = b$ e de amplitude $= gg$; cuja base inferior do tubo seria perfurada com uma abertura $CC = kk$, através da qual a água eflua. Seja também $\frac{k^4}{f^4} = \mathfrak{f} = 0$, e tenham sido estabelecidos

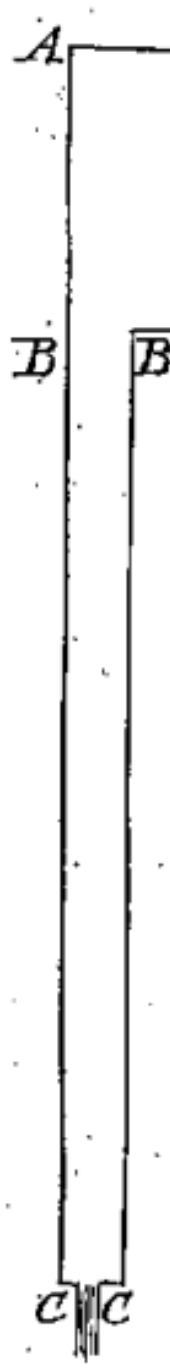
Fig.5.

$$\mathfrak{A} = e^{\frac{-\propto a}{f}}, \quad \mathfrak{B} = e^{\frac{-\propto b}{g}}, \quad A = \frac{1-\mathfrak{A}}{\propto} \quad \text{e} \quad B = \frac{1-\mathfrak{B}}{\propto}$$

Diante de $\cos\zeta = 1$ e $\mathfrak{g} = \frac{k^4}{g^4}$ será a velocidade, com que a água eflui através do orifício $CC$, devida à altitude $v$, que seria

$$v = \frac{A\mathfrak{B}f + Bg - l(1 + \mathfrak{A}\mathfrak{B})}{1 - \propto B\mathfrak{g}}.$$

46. E assim para valores tão próximos aos reais tanto quanto desejemos, teremos

$$v\left(1 - \frac{\alpha b k^4}{g^5}\left(1 - \frac{\alpha b}{2g} + \frac{\alpha^2 b^2}{6g^2} - \frac{\alpha^3 b^3}{24g^3} + etc.\right)\right)$$

$$= a + b - \alpha a\left(\frac{a}{2f} + \frac{b}{g}\right) + \alpha^2 a\left(\frac{aa}{6ff} + \frac{ab}{2fg} + \frac{bb}{2gg}\right)$$

$$-\alpha^3 a\left(\frac{a^3}{24f^3} + \frac{a^2 b}{6ffg} + \frac{ab^2}{4fgg} + \frac{b^3}{6g^3}\right) + etc.$$

$$-\frac{\alpha bb}{2g} + \frac{\alpha\alpha b^3}{6gg} - \frac{\alpha^3 b^4}{24g^3} + etc.$$

$$-\alpha l\left(\frac{a}{f} + \frac{b}{g}\right) + \frac{1}{2}\alpha^2 l\left(\frac{a}{f} + \frac{b}{g}\right)^2 - \frac{1}{6}\alpha^3 l\left(\frac{a}{f} + \frac{b}{g}\right)^3 + etc.,$$

que é reduzida à seguinte forma:

$$v\left(1-\frac{k^4}{g^4}\left(\frac{\alpha b}{g}-\frac{\alpha^2 b^2}{2g^2}+\frac{\alpha^3 b^3}{6g^3}-etc.\right)\right)$$

$$= a+b-\alpha\left(\frac{aa}{2f}+\frac{ab}{g}+\frac{bb}{2g}+\frac{al}{f}+\frac{bl}{g}\right)$$

$$+\alpha\alpha\left(\frac{a^3}{6ff}+\frac{aab}{6fg}+\frac{abb}{2gg}+\frac{b^3}{6gg}+\frac{1}{2}l\left(\frac{a}{f}+\frac{b}{g}\right)^2\right)$$

$$-\alpha^3\left(\frac{a^4}{24f^3}+\frac{a^3 b}{6ffg}+\frac{aabb}{4fgg}+\frac{ab^3}{6g^3}+\frac{b^4}{24g^3}+\frac{1}{6}l\left(\frac{a}{f}+\frac{b}{g}\right)^3\right)+etc.$$

47. Se a altitude do vaso supremo $AB = a$ for muito curta, e, além disso, a amplitude $ff$ muito maior que a amplitude do tubo $gg$, e simultaneamente o comprimento deste tubo $BC = b$ seja suficientemente grande, fica evidente que a fração $\frac{b}{g}$ é extremamente superior à fração $\frac{a}{f}$; desta forma consequentemente perante aquela desprezada, será

$$v\left(1-\frac{k^4}{g^4}\left(\frac{\alpha b}{g}-\frac{\alpha^2 b^2}{2gg}+\frac{\alpha^3 b^3}{6g^3}-etc.\right)\right)$$

$$= a+b-\frac{\alpha b}{g}\left(a+\frac{1}{2}b+l\right)+\frac{\alpha\alpha bb}{gg}\left(\frac{1}{2}a+\frac{1}{6}b+\frac{1}{2}l\right)$$

$$-\frac{\alpha^3 b^3}{g^3}\left(\frac{1}{6}a+\frac{1}{24}b+\frac{1}{6}l\right)+etc.$$

e esta série converge tanto mais, quanto menor for a fração $\frac{\alpha b}{g}$.

48. No entanto, por causa de certos consistentes experimentos na medida em que foi possível coletar de passagem, o valor do próprio $\alpha$ surgiu aproximadamente $= \frac{1}{4000}$. Logo, contanto que o número $\frac{b}{g}$ será menor que $4000$, bastará que no limite subsistam os termos mais importantes, assim que seria

$$v\left(1-\frac{\alpha b}{g}\cdot\frac{k^4}{g^4}\right)=a+b-\frac{\alpha b}{g}\left(a+\frac{1}{2}b+l\right)$$

ou

$$v = a + b - \frac{\alpha b}{g}\left(a + \frac{1}{2}b + l - \frac{k^4}{g^4}(a + b)\right).$$

49. Se o orifício $CC = kk$ for mínimo, de tal modo que $\frac{k^4}{g^4}$ por zero possa ser tomado, será

$$v = a + b - \frac{\alpha b}{g}\left(a + \frac{1}{2}b + l\right),$$

mas se no entanto, o tubo inferior esteja livre ou $gg = kk$, será considerado

$$v = a + b - \frac{\alpha b}{g}\left(l - \frac{1}{2}b\right).$$

Portanto, neste caso a velocidade do efluxo é mínima e é impedida ao máximo pela fricção; no entanto o caso é visto poder ocorrer, de maneira que se torna em $v = a + b$, como se a fricção não estivesse presente.

50. Certamente, se por acaso devesse acontecer, $b = 2l$, ou sendo $l$ de trinta pés, fosse $b = 60$ pés: e ainda de tal modo, se fosse $b > 60$ pés, o nosso cálculo fornecesse $v > a + b$. Entretanto ainda que certíssimo é, a fricção nesta situação não pode ser de maneira que a água exile com maior velocidade, como se não estivesse presente a fricção. Declaro, pois, que casos deste tipo certamente não possam aqui ocorrer, pelo motivo que a água não adira em toda a parte do tubo, mas deixasse para trás o vácuo, posto que sem dúvida o cálculo não possa ser a eles aplicado.

51. Para que isto seja visto mais claramente, deve ser observado que a água enquanto preencher toda a cavidade do tubo, até o ponto que é oprimida contra sua parede: sem que portanto ceda, para que a pressão da água dentro do tubo em algum lugar ou desapareça ou se faça um tanto negativa, de fato nesse lugar a parede do tubo abandonará e deixará para trás o vácuo, posto que sem dúvida o movimento da água que sem contestação há-de existir de outra forma, e é definido através do cálculo. Efetivamente o cálculo não é capaz de valer neste local, a não ser que o estado de compressão da água no tubo neste local seja positivo.

52. Coloquemos, portanto, o estado de compressão da água na seção $BB$ sendo $=\ R$, e por causa do parágrafo 36 teremos

$$l = Re^{\frac{-\alpha b}{g}} + \frac{g}{\alpha}\left(1 - e^{\frac{-\propto b}{g}}\right)\cos\theta + \frac{k^4}{g^4}v - v$$

e por esta razão

$$R = e^{\frac{\alpha b}{g}} l - \frac{g}{\alpha}\left(e^{\frac{\propto b}{g}} - 1\right) + v\left(1 - \frac{k^4}{g^4}\right)e^{\frac{\alpha b}{g}}\,.$$

Já por causa de

$$e^{\frac{\alpha b}{g}} = 1 + \frac{\alpha b}{g} + \frac{\alpha^2 b^2}{2g^2}\,,$$

será

$$R = l + \frac{\alpha b l}{g} - b - \frac{\alpha b b}{2g} + \left(1 + \frac{\alpha b}{g}\right)v\left(1 - \frac{k^4}{g^4}\right)$$

e sendo substituído pelo valor de $v$ descoberto antes, resultará desprezados os termos multiplicados por $\alpha$

$$R = l + a - \frac{k^4}{g^4}(a + b)\,.$$

53. A menos que seja

$$l + a > \frac{k^4}{g^4}(a + b)\,,$$

hipótese assumida no cálculo que não pode ter lugar; e por este motivo deve o comprimento do tubo $b$ ser menor, que

$$\frac{g^4}{k^4}(l + a) - a\,.$$

Razão pela qual se o tubo inferior $BC$ esteja totalmente aberto ou $gg = kk$, é necessário, que seja $b < l$: donde a hipótese é arruinada, se fosse $b = 2l$. Caso $gg = kk$, o comprimento do tubo $BC = b$ não pode superar trinta pés, donde colocando $b = l$ para a máxima velocidade do efluxo, ou que é mínima impedida pela fricção, que é devida à altitude

$$v = a + b - \frac{\alpha ll}{2g}$$

e assim é menor, que aquela caso não existisse a fricção.

54. Mas em geral, caso seja $\frac{a}{f}$ tão pequeno, que perante $\frac{b}{g}$ possa ser considerado zero, será

$$\mathfrak{A} = 1 \quad , \qquad A = \frac{a}{f} \; ;$$

$$v = \frac{ae^{\frac{-\propto b}{g}} + \frac{g}{\propto}\left(1 - e^{\frac{-\propto b}{g}}\right) - l\left(1 - e^{\frac{-\propto b}{g}}\right)}{1 - \frac{k^4}{g^4}\left(1 - e^{\frac{-\propto b}{g}}\right)},$$

donde, caso o tubo inferior $BC$ esteja totalmente aberto ou $\frac{k^4}{g^4} = 1$, será

$$v = a + \frac{g}{\propto}\left(e^{\frac{\propto b}{g}} - 1\right) - l\left(e^{\frac{\propto b}{g}} - 1\right).$$

55. Se pois o tubo $BC$ for tão fino, tal que seja $g = \alpha l$, a água em $CC$ não efluirá com maior velocidade, que efluisse com o tubo em $BB$ livre. E ainda se, além disso, o tubo $BC$ for mais fino ou $g < \propto l$, por exemplo $g < \frac{1}{2} \propto l$, será feita

$$v = a + \frac{1}{2} l\left(e^{\frac{2b}{l}} - 1\right) \; ,$$

e se a altura $b$ for muito pequena perante $l$, será

$$v = a - b - \frac{bb}{l} \; ;$$

por isso se não for $a > b + \frac{bb}{l}$ , a água simplesmente não efluirá, isto é, nem que seja $a > b$.

56. Então o tubo $BC$ pode ser tão fino, para que a fricção retenha completamente o efluxo da água através de seu orifício $CC$, e logo muito comodamente o valor da letra $\alpha$ através de experimentos poderá ser investigado. Seja assim fixado a um vaso suficientemente amplo $AB$, no qual a água ocupa uma altitude $AB = a$ não demasiadamente grande, um tubo, que costuma ser chamado, capilar, que existindo com amplitude $= gg$, seja estabelecido $g = \frac{1}{n} \propto l$, e sendo

$$v = a - \frac{(n-1)}{n} l \left(e^{\frac{nb}{l}} - 1\right) = a - (n-1)b \ ,$$

instalo este tubo que fosse no início tão longo, para que nenhuma água efluísse; imediatamente depois, seja diminuído de modo sucessivo o comprimento deste, até que a água comece a escoar: com este comprimento anotado, qual seja $= b$, sendo $a - (n-1)b = 0$, será $n = \frac{a+b}{b}$ e por esta razão

$$\propto = \frac{a+b}{b} \cdot \frac{g}{l} .$$

57. O mesmo experimento também deste modo pode ser montado, para que um determinado comprimento do tubo capilar $BC$ seja adotado, qualquer que seja o tamanho, tal que, se um mínimo de água no vaso superior seja derramada, nada eflua; então certamente seja aumentada a altura da água no vaso até o ponto, em que a água através do tubinho capilar inicie a efluir. Estando isto acontecendo, seja anotada a altitude $= a$, e porque é colocado o comprimento do tubinho $b$ bem como já tendo sido conhecida sua amplitude $gg$, conforme antes é obtido

$$\propto = \frac{a+b}{b} \cdot \frac{g}{l} .$$

58. Também evidente é o tubinho, que queremos usar, que deve ser tão estreito, para que seja $g < \propto l$; pois se fosse $g > \propto l$, o uso deste caso nunca servirá, para que o efluxo da água seja estorvado. Mas seja então que fosse permitido supor, $\propto = \frac{1}{4000}$, por causa de $l = 30\ pés$ deveria ser

$$g < \frac{3}{400}\ pés \qquad \text{ou} \qquad g < 0{,}0075\ pés$$

assim o diâmetro do tubinho deveria ser menor, que $\frac{85}{10000}$ $pés$ ou que $\frac{1}{118}$ $pés$. Então o limite, inferior ao qual o diâmetro do tubinho devesse ser tomado, fosse em torno de uma só linha.

59. Se fosse $g = \propto l$ ou $g < \propto l$ e $a = 0$, então se um tubo deste tipo, de qualquer comprimento que fosse, que com água seja enchido, nada do interior dele escoará, mesmo que seja mantido na vertical. Então, sendo $l$ um comprimento de 30 pés, facilmente poderá ser investigado o valor da letra $\propto$ através de experimentos. Isto é, sejam preparados diversos tubinhos, dos quais as amplitudes $gg$ sejam diferentes, que estejam os mesmos repletos com água; então dos mais amplos certamente a água efluirá, sem dúvida doutro modo dos mais estreitos, então tudo que é necessário, que deste tubinho, que abundante água não sai, seja medida a amplitude $gg$, e será então $\propto = \frac{g}{l}$. Mas este verdadeiro efluxo deve ser observado, com o qual a água eflui estendida continuamente; pois se somente cai gota-a-gota, será o mesmo como se a água não efluisse. Pois finalmente tendo certamente discernida a fricção neste movimento, até que a água se acalme, devido à gravidade as gotas são gradualmente arrancadas e cairão, o que a fricção não tem força de impedir.

60. Fica, portanto, patente a magreza desses tipos de tubinhos, que mesmo que sejam abertos de ambos os lados, não deixam efluir a água contida, que pende pelo peso da atmosfera $l$, visto que isto acontece, se $g < \propto l$ . Devido a isto, observamos no vácuo, onde $l = 0$, que estes tubinhos eliminam totalmente a água, a menos até o ponto devido à natural constituição, que é particular aos tubinhos capilares, a água neles fique retida até certa altitude, mas que efluirá a água remanescente no vácuo, que na atmosfera caia toda gota-a-gota.

61. Também geralmente é entendido que devido à pressão da atmosfera que o efluxo da água é sempre retardado, e que certamente a fricção é a causa desta retardação: com efeito, vimos (54), que quanto menor seja a pressão da atmosfera $l$, por isto com maior velocidade a água jorra. Mas se, portanto, seja a pressão da atmosfera diminuída não importa porque causa, a água mais velozmente eflui do vaso; que por esta razão suponho ser, que o efluxo da água seja considerado ser acelerado pela eletricidade.

62. Não posso desprezar este fenômeno, aliás fácil de ser explicado, que certamente, se o vaso superior seja infinitamente estreito, ou se sua amplitude $ff$ desapareça completamente, de tal modo que o tubo $BC$ acima esteja inteiramente fechado, a água não eflui através do orifício $CC$, a não ser que o tubo fosse muito longo. De fato este fenômeno, desprezando a fricção, esta teoria não vai ao encalço, não se anulando o próprio valor de $v$, mesmo que seja colocado $f = 0$.

63. Mas levando em conta a fricção no duto, como, por causa de $f = 0$, torne-se $\mathfrak{A} = 0$, será da fração, cujo valor fornece o próprio $v$, o numerador

$$= Bg - l = \left(1 - e^{\frac{-\propto b}{g}}\right)\frac{g}{\alpha} - l\,,$$

efetivamente para o denominador não é trabalho que me ocupe, que envolvesse a fração $\frac{k^4}{f^4} = \mathfrak{f}$, a qual desprezamos no parágrafo 45. Mas do numerador é reconhecido como sempre

$$1 - e^{\frac{-\propto b}{g}} < \frac{\propto l}{g}\,,$$

poder haver aqui um efluxo nulo. Este, portanto, acontece se

$$e^{\frac{-\propto b}{g}} > 1 - \frac{\propto l}{g}$$

ou

$$= 1 - \frac{\propto b}{g} + \frac{\propto^2 b^2}{2g^2} - \frac{\propto^3 b^3}{6g^3} + etc. > 1 - \frac{\propto l}{g}$$

e por esta razão se

$$l > b - \frac{\propto bb}{2g} + \frac{\propto^2 b^3}{6gg} - etc.$$

Ou a água não eflui, todas as vezes que for

$$\frac{\propto b}{g} < -log\left(1 - \frac{\propto l}{g}\right)$$

ou

$$b < l - \frac{\propto ll}{2g} + \frac{\propto^2 l^3}{3gg} - \frac{\propto^3 l^4}{4g^3} + etc.$$

Portanto, a não ser que o comprimento do tubo $b$ fosse maior que este, a água não é capaz de efluir.

64. Além disso, percebamos diante de algum caso determinado, quanto seria o efeito da fricção, colocado o valor $\propto = \frac{1}{4000}$, que na verdade é visto não ser muito inconsistente. Seja então a altitude do amplíssimo vaso supremo $AB = a = \frac{1}{3}\ p\acute{e}$, o comprimento do tubo anexo $BC = b = 4\ p\acute{e}s$, sua amplitude $gg = \frac{1}{2500}\ p\acute{e}$ ou $g = \frac{1}{50}\ p\acute{e}$ este tubo embaixo seja completamente aberto, posto que se torne $kk = gg$; daqui será

$$\frac{\propto b}{g} = \frac{1}{20} \qquad e \qquad e^{\frac{\propto b}{g}} - 1 = 0{,}05127$$

e ainda $\frac{g}{\propto} - l = 50$.

$$v = \frac{1}{3} + 2{,}564 = 2{,}897 \text{ pés}$$ [12].

Portanto, para este caso, a altitude que é devida pela velocidade que a água eflui, é tão grande quanto 2,897 pés, com a fricção removida $= 4\frac{1}{3}\ p\acute{e}s$.

65. Se o efeito da fricção fosse menor, que aquele que assumi, a velocidade da água se aproximasse mais em direção à altitude de $4\frac{1}{3}\ p\acute{e}s$. Para que isto apareça mais claramente, será

$$\frac{\propto b}{g} = \frac{1}{30} \qquad e \qquad e^{\frac{\propto b}{g}} - 1 = 0{,}033894$$

[12] Edição primeira: $v = \frac{1}{3} + 2{,}563 = 2{,}896$. Valor mais acurado retirado da fórmula do parágrafo 64: 3,028. C.T. Corrigiu

e ainda $\frac{g}{\propto} - l = 50$, de onde será feito

$$v = \frac{1}{3} + 3{,}05056 = 3{,}8389\ \text{pés}^{13}.$$

Portanto a altitude que é devida pela velocidade havesse de ser agora 3,8370, e por isto se aproximasse mais perto de $4\frac{1}{3}\ pés$.

66. Mas com um vaso construído dessa forma, pode ser investigado através de um experimento, com quanta quantidade de água eflui em certo tempo. Pois, deixemos efluir num primeiro minuto uma quantidade de água, qual seja $m$ pés cúbicos, e que igualmente em pés será $v = 100000\, m\, m$, cujo valor seja comparado com aqueles, os quais determinamos para $v$ a partir das hipóteses $\propto = \frac{1}{4000}$ e $\propto = \frac{1}{6000}$, ou se estas hipóteses menos concordem, sejam desenvolvidos outros cálculos, assim será conhecido o verdadeiro valor do próprio $\propto$. Conduzirá também diversos experimentos que dessa forma estabeleci, até que sejamos capazes de estar plenamente certos acerca do verdadeiro valor do próprio $\propto$.

## CASO II
## SE A ÁGUA DO VASO SUPREMO EFLUI ATRAVÉS DE UM TUBO CILÍNDRICO VERTICAL COMPOSTO DE DUAS PARTES CILÍNDRICAS

67. Seja (Fig. 6) a amplitude do vaso supremo $AA = ff$, que seja muito grande, e que sua altura $AB = a$, de maneira que estabeleço que o vaso é continuamente conservado cheio com água. Que a partir deste vaso a água flua no tubo $BC$, de cuja amplitude seja $BB = gg$ e altitude $BC = b$: seja, além disso, anexado àquele tubo em $CC$ um tubo vertical $CD$, de amplitude $= hh$ e de comprimento $CD = c$: de cuja base inferior tenha sido perfurada uma abertura $DD = kk$, através da qual a água eflua na atmosfera. Que também $l$ expresse a altitude da coluna d'água de mesmo peso da atmosfera: seja também $kk$ perante $ff$ tão pequeno, para que $\mathfrak{f} = \frac{k^4}{f^4}$ possa ser tomado como zero.

---

[13] Edição primeira: $v = \frac{1}{3} + 3{,}05046 = 3{,}38379$. Valor mais acurado retirado da fórmula do parágrafo 64: 3,48729. C.T. Corrigiu

68. Sendo estes estabelecidos por causa da brevidade

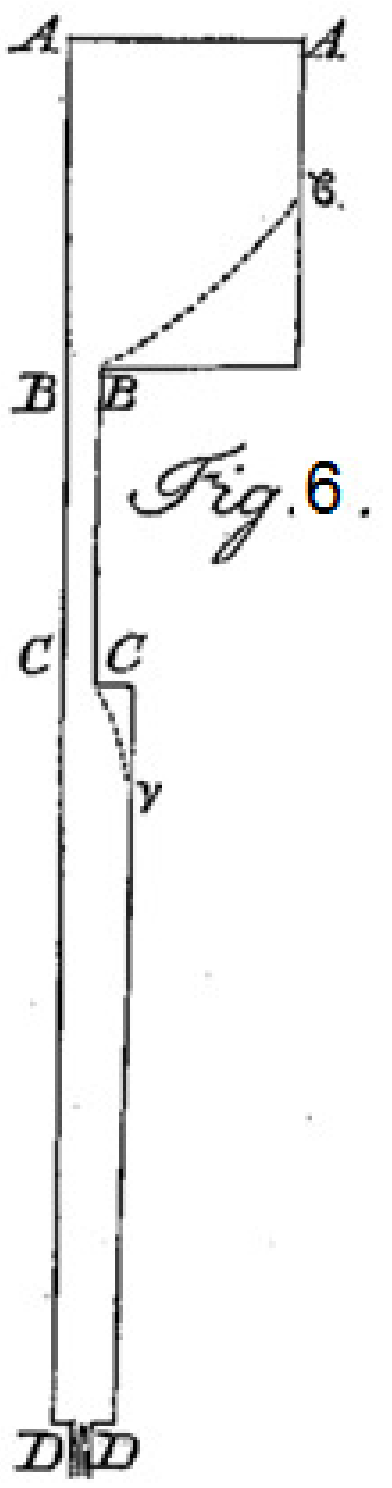

Fig. 6.

$$\mathfrak{A} = e^{\frac{-\propto a}{f}}, \quad A = \frac{1-\mathfrak{A}}{\propto}, \quad \mathfrak{B} = e^{\frac{-\propto b}{g}}, \quad B = \frac{1-\mathfrak{B}}{\propto},$$

$$\mathfrak{C} = e^{\frac{-\propto c}{h}}, \quad C = \frac{1-\mathfrak{C}}{\propto},$$

além disso, certamente

$$\mathfrak{g} = \frac{k^4}{g^4} \qquad e \qquad \mathfrak{h} = \frac{k^4}{h^4}.$$

Agora a água que eflui através do orifício $DD = kk$ é descoberta com a velocidade devida pela altitude, qual seja

$$v = \frac{A\mathfrak{B}\mathfrak{C}f + B\mathfrak{C}g + Ch - (1-\mathfrak{A}\mathfrak{B}\mathfrak{C})l}{1-(1-\mathfrak{B})\mathfrak{C}\mathfrak{g} - (1-\mathfrak{C})\mathfrak{h}},$$

donde, como já acima apontamos, resulta, caso a fricção desapareça

$$v = a + b + c,$$

desse modo a altitude que fosse devida pela velocidade iguala à toda altitude $AD$.

69. Se a amplitude do tubo do meio $BC$ fosse igual à amplitude do mais baixo $CD$, voltaria para o caso que precede; mas se, no entanto fosse mais amplo, como pudesse ser visto como uma parte do vaso supremo, e pura e simplesmente haveria voltado para o primeiro caso. Por isso, portanto, se os fenômenos, que são próprios ao caso, desejamos descobrir, o tubo médio $BC$ virá juntamente numa posição vertical, muito mais estreito que o vaso superior ou que o tubo mais baixo $CD$. Está confirmado pelos experimentos, que por causa da fineza deste tipo de passagem média $BC$, a velocidade do efluxo não é apreciavelmente diminuída, donde a causa desta diminuição é abertamente atribuída à fricção, com a fricção removida a amplitude deste tubo $gg$ não afetaria a velocidade do efluxo.

70. Contudo, se a amplitude do tubo $BC$ é muito menor, que a do superior e do inferior, é evidente que o veio de água, que do vaso superior para ele entra, imediatamente é contraído, e com modo similar, sai daí no tubo mais amplo $CD$, mas também depois de egresso permanece mais estreito; e deste modo a água desta forma será movida, como se passasse através do tubo $\beta BBCC\gamma$, ainda que deva tomar-se em conta que tanto o vaso superior quanto o tubo inferior contrair-se de qualquer maneira. Esta circunstância será considerada no cálculo, se o comprimento do tubo mais fino $BC$ um tanto maior seja considerado, que é na realidade, até o ponto que seja subtraída a altura dos tubos contíguos.

71. Assumirei, portanto, que é feita, neste momento, esta alteração na designação das alturas $a, b, c$, para que seja a quantidade $b$ tanto maior, quanto sejam verdadeiramente menores $a$ e $c$, como na realidade são tomados. Desta forma, portanto, a altura $b$ é considerada tanto maior, quanto menor tenha sido a amplitude deste tubo, diante da amplitude tanto do superior quanto do inferior. Por isso, se o tubo $BC$ tenha sido estreitíssimo ou $gg$ mínimo, e também a altura deste $BC$ fosse mínima, contudo o valor da letra $b$ deverá ser considerado extraordinário; e se $BC$ quase desaparecesse, o que acontece, se o fundo do vaso superior fosse perfurado com uma abertura exígua, todavia a letra $b$ caberia no cálculo um valor pequeno.

72. Seja, pois, a amplitude quase evanescente do tubinho $BC$ ou $g$ quase igual a zero, de tal forma que o valor do próprio $b$ alcance uma magnitude moderada, ainda que acidentalmente a própria altura $BC$ seja mínima; e será $\frac{b}{g}$ um número assaz grande, pelo qual o próprio $\mathfrak{B} = e^{\frac{-\alpha b}{g}}$ resultará numa fração muito menor que a unidade, tal que, se fosse $g = 0$, terá sido integralmente $\mathfrak{B} = 0$. Além disso, verdadeiramente, tornar-se-á $\mathfrak{g} = \frac{k^4}{g^4}$ uma quantidade máxima. Visto que verdadeiramente as amplitudes $ff$ e $hh$ são estabelecidas não assaz pequenas, será

$$\mathfrak{A} = 1 - \frac{\alpha a}{f} + \frac{\alpha\alpha aa}{2ff}, \qquad \mathfrak{C} = 1 - \frac{\alpha c}{h} + \frac{\alpha^2 cc}{2hh}$$

e

$$A = \frac{a}{f} - \frac{\alpha aa}{2ff} \qquad e \qquad C = \frac{c}{h} - {}^{14}\frac{\alpha cc}{2hh}\,.$$

73. No entanto, antes que possamos definir o próprio efluxo, deve ser observado, que se a água neste vaso permanece contínua e é capaz de aderir-se nas laterais do vaso; o que procuramos ao final é o estado de compressão da água em $CC$, que seria expresso pela altitude $R$, e será

$$l = R\mathfrak{C} + Ch + \mathfrak{h}v - v\,,$$

daí que é achado:

$$R = \frac{l - Ch + (1 - \mathfrak{h})v}{\mathfrak{C}}\,,$$

que se a magnitude fosse negativa, a água não permaneceria continua e por esta razão seria adverso o cálculo do efluxo.

74. Então sendo estabelecido $\mathfrak{B}$ muito pequeno e por isso $B = \frac{1}{\alpha}$

$$v = \frac{a\mathfrak{B}\mathfrak{C} + \frac{g}{\alpha}\mathfrak{C} + Ch - l}{1 - \mathfrak{C}\mathfrak{g} - \alpha C\mathfrak{h}} = \frac{\frac{g}{\alpha}\mathfrak{C} + Ch - l}{1 - \mathfrak{C}\mathfrak{g}}\,,$$

com os menores termos desprezados será

$$R = \frac{l - Ch}{\mathfrak{C}} + \frac{(1 - \mathfrak{h})\frac{g}{\alpha}}{1 - \mathfrak{C}\mathfrak{g}} - \frac{(1 - \mathfrak{h})(l - Ch)}{\mathfrak{C}(1 - \mathfrak{C}\mathfrak{g})},$$

$$R = \frac{(1 - \mathfrak{h})\frac{g}{\alpha}}{1 - \mathfrak{C}\mathfrak{g}} - \frac{(\mathfrak{C}\mathfrak{g} - \mathfrak{h})(l - Ch)}{\mathfrak{C}(1 - \mathfrak{C}\mathfrak{g})}\,.$$

Mas sendo muito próximo

$$\mathfrak{C} = 1 \qquad e \qquad Ch = c\,,$$

será

$$v = \frac{\frac{g}{\alpha} + c - l}{l - \mathfrak{g}} \qquad e \qquad R = \frac{\frac{g}{\alpha}(1 - \mathfrak{h}) - (\mathfrak{g} - \mathfrak{h})(l - c)}{1 - \mathfrak{g}}\,.$$

[14] Edição primeira: $+$. Tradutor Corrigiu

75. Portanto, fica logo patente, se fosse $\mathfrak{g} = 1$ ou $\mathfrak{g} > 1$, que o efluxo jamais será conduzido ao estado de uniformidade e por esta razão o movimento a partir de fórmulas, as quais são apropriadas a este estado, não pode ser definido inteiramente; por isso sem se mexer desde o início do cálculo que de tal modo devesse ser preparado, para que a altura $v$ como variável fosse introduzida. Mas se, no entanto, tivesse sido $\mathfrak{g} < 1$ ou $kk < gg$, o efluxo certamente passará a uniforme e a água efluirá, se tivesse sido

$$\frac{g}{\alpha} + c > l \quad \text{e simultaneamente} \quad \frac{g}{\alpha}(1 - \mathfrak{h}) > (\mathfrak{g} - \mathfrak{h})(l - c);$$

mas se, no entanto, seja

$$\frac{g}{\alpha} + c = l \qquad \text{ou} \qquad \frac{g}{\alpha} + c < l\,,$$

a água simplesmente não efluirá, ainda que assumimos existente $kk < gg$. Todavia caso seja

$$\frac{g}{\alpha} + c > l \qquad \text{por exemplo} \qquad \frac{g}{\alpha} + c = l + \gamma,$$

o efluxo alcançará a uniformidade, todas as vezes que $c < l$. Mas nos casos, para os quais $c > l$, este ocorrerá com muito mais razão, pois, diante de $\mathfrak{h} < g$, então o valor do próprio $R$ sempre é positivo.

## CASO III
## SE A ÁGUA DO VASO CONSTANTEMENTE CHEIO EFLUI ATRAVÉS DE UM TUBO HORIZONTAL

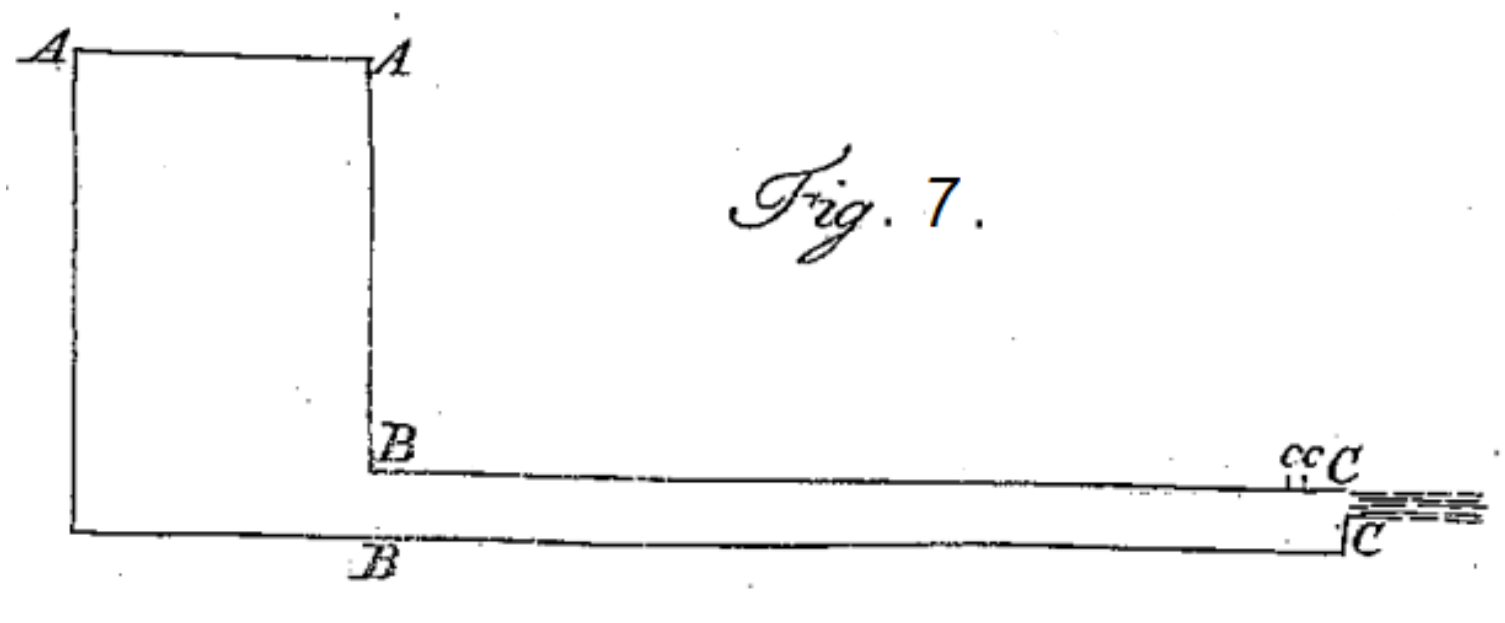

76. Seja (Fig. 7) a amplitude do vaso $AA = ff$ e altura $AB = a$, verdadeiramente de um tubo fixado horizontalmente o comprimento $BC = b$, a amplitude $BB = gg$, e um lume através do qual a água eflui, $CC = kk$. Seja colocado por causa da brevidade

$$\mathfrak{A} = e^{\frac{-\propto a}{f}}, \quad A = \frac{1-\mathfrak{A}}{\propto}, \quad \mathfrak{B} = e^{\frac{-\propto b}{g}},$$

$$B = \frac{1-\mathfrak{B}}{\propto}, \quad \mathfrak{f} = \frac{k^4}{f^4} = \mathfrak{f} \quad \text{e} \quad \mathfrak{g} = \frac{k^4}{g^4} \;;$$

será a altura devida pela velocidade do efluxo, diante de $\zeta = 90º$ e $\cos\zeta = 0$, denotando $l$ a altitude de 30 pés,

$$v = \frac{A\mathfrak{B}f - (1-\mathfrak{A}\mathfrak{B})l}{1 - \mathfrak{B}\mathfrak{f} - \alpha B\mathfrak{g}}.$$

77. Se o vaso $AB$ fosse muito grande, seria

$$\mathfrak{f} = 0, \quad \mathfrak{A} = 1 - \frac{\alpha a}{f} \quad \text{e} \quad A = \frac{a}{f} - \frac{\alpha a^2}{2ff},$$

donde a velocidade do efluxo será devida à altitude

$$v = \frac{a\mathfrak{B}\left(1 - \frac{\alpha a}{2f}\right) - \left(1 - \mathfrak{B} + \frac{\alpha a}{f}\,\mathfrak{B}\right)l}{1 - (1-\mathfrak{B})\mathfrak{g}},$$

donde com $\mathfrak{g} = \frac{k^4}{g^4}$, que não pode superar a unidade, o denominador será sempre uma quantidade positiva $> Bg$: cuja evidência é que certamente o efluxo da água chegará ao estado de uniformidade.

78. Também, para que o movimento eflua a água, é necessário, que seja

$$a\mathfrak{B}\left(1 - \frac{\alpha a}{2f}\right) > \left(1 - \mathfrak{B} + \frac{\alpha a}{f}\,\mathfrak{B}\right)l$$

ou

$$\mathfrak{B} > \frac{fl}{f(a+l) - \alpha a\left(\frac{1}{2}a + l\right)} \;;$$

e por esta razão

$$e^{\frac{\propto b}{g}} < 1 + \frac{a}{l} - \frac{\alpha a a}{2fl} - \frac{\alpha a}{f} \;;$$

daqui com a tomada dos logaritmos seja necessário

$$\frac{\propto b}{g} < log\left(1+\frac{a}{l}\right) - \frac{\alpha a(a+2l)}{2f(a+l)},$$

onde $log\left(1+\frac{a}{l}\right)$ denota o logaritmo hiperbólico do número $1+\frac{a}{l}$. Portanto, o efluxo cessará, caso seja

$$b > \frac{g}{\propto} log\left(1+\frac{a}{l}\right) - \frac{ag(a+2l)}{2f(a+l)}.$$

79. Por isso, quanto mais longo fosse o tubo horizontal $BC$, tanto mais lentamente a água efluiria, e o seu comprimento pode crescer de tal maneira que a água através dele simplesmente não eflua; que naturalmente ocorrerá se fosse

$$b > \frac{g}{\propto} log\left(1+\frac{a}{l}\right) - \frac{ag(a+2l)}{2f(a+l)}.$$

Isto também deve ser percebido, se fosse feita uma abertura $CC$ na parte de cima da extremidade do tubo $BC$. Pois certamente efluiria se fosse na parte de baixo, por conta da própria gravidade da água no tubo horizontal, a qual não foi por mim contemplada no cálculo.

80. Se a altura $a$ for menor que $l$, será aproximadamente

$$log\left(1+\frac{a}{l}\right) = \frac{a}{l}.$$

Portanto, assim que o comprimento do tubo $b$ terá superado esta quantidade

$$\frac{ga}{\propto l} - \frac{ag(a+2l)}{2f(a+l)},$$

o efluxo de água através do orifício $CC$ cessará. E porque o segundo termo é mínimo com relação ao primeiro, a água não efluirá abundante, quando terá sido $b > \frac{ga}{\propto l}$. Logo seja necessário para que a água eflua,

$$b < \frac{ga}{\propto l}.$$

81. Portanto, se a altura do vaso $AB = a$ for de um único pé, diante de $l = 30\ pés$, o efluxo da água será retido se o comprimento do tubo $BC$ fosse maior que $\frac{g}{30\propto}$. Logo, um novo modo é obtido para ser determinado o valor da letra $\propto$: sendo através de experimentos explorado o comprimento do tubo horizontal $b$, cuja amplitude $gg$ seja anotada, quando cessa o efluxo, será

$$\propto = \frac{ga}{bl}.$$

82. Portanto experimentos deste tipo poderão ser instituídos com tubos não tanto estreitos, da mesma maneira que com o modo antes exposto, donde este mesmo modo é reconhecido ter sido antes aplicado. Pois, tubos muito estreitos, que costumam ser chamados capilares, sejam providos com propriedades singulares, sempre uma dúvida seria deixada para trás, se, perante estas propriedades, o efluxo da água através destes tipos de tubos não sofra uma perturbação particular, de modo que o valor do próprio $\propto$ daí coletado fosse dado como duvidoso.

83. A fim de que eu apresente um exemplo de experimentos deste tipo, estabeleçamos que exista um vaso amplíssimo $AAB$, que exista o comprimento $b = 2$ pés e a amplitude $gg = \frac{1}{625}$ pés quadrados, e por esta razão $g = \frac{1}{25}$ pés, de um tubo $BC$ verdadeiramente horizontal; também o lume dele $kk$ que de tal forma seja pequeno, para que $\mathfrak{g} = \frac{k^4}{g^4}$ possa ser considerado como zero. Daqui, perante o vaso amplíssimo $AAB$ poderá ser ignorada a fração $\frac{\alpha a}{f}$, e será considerado

$$\mathfrak{B} = e^{\frac{-\propto b}{g}} = 1 - \frac{1}{80},$$

tomado $\propto = \frac{1}{4000}$. Certamente será obtido

$$v = a - \frac{a - 30}{80} = \frac{79a - 30}{80}.$$

Para que então neste caso a água eflua com ação, é necessário, que seja

$$a > \frac{30}{79} \text{ pés.}$$

84. Se tivéssemos colocado $\propto = \frac{1}{6000}$, surgiria

$$\mathfrak{B} = 1 - \frac{1}{120} \quad \text{e} \quad v = a - \frac{a - 30}{120} = \frac{119a - 30}{120},$$

logo, a água começaria a efluir imediatamente certamente quando a altura da água $a$ no vaso $AB$ superasse $\frac{30}{119}$ pés. Então, com $\propto$ colocado incógnita seria

$$\mathfrak{B} = 1 - 50 \propto \quad \text{e} \quad v = a(1 - 50 \propto) - 1500 \propto,$$

seria aumentada gradualmente a altura da água no vaso $AB$, até que a água através do orifício $CC$ inicie a efluir, e seria anotada então a altura $AB = a$ em pés, será

$$\propto = \frac{a}{50a + 1500}.$$

85. Se o mesmo experimento fosse montado com outro tubo horizontal, do qual tanto o comprimento $b$ quanto a amplitude $gg$ seja qualquer, mas, entretanto, de tal maneira que $\frac{\alpha b}{g}$ permaneça uma fração assaz pequena, e que seja suficientemente exata

$$\mathfrak{B} = 1 - \frac{\alpha b}{g} \quad \text{e} \quad v = a\left(1 - \frac{\alpha b}{g}\right) - \frac{\alpha b l}{g},$$

e ainda que a altura da água no vaso $AB$ seja anotada, quando a água através do orifício $CC$ inicie a efluir pela primeira vez, assim será obtido

$$\alpha = \frac{ag}{b(a + l)}$$

e que é visto como um certíssimo modo de investigar o verdadeiro valor do próprio $\alpha$.

86. No entanto, consideremos como conhecido o valor do próprio $\alpha$, qual seja $\propto = \frac{1}{4000}$, e vejamos por um exemplo, quanta debilitação deva resultar por causa da fricção nas fontes com jorros d'água. Seja então o vaso $AB$ assaz alto, qual seja $a = 100$ pés e sua amplitude $ff = 1$ pé quadrado ou $f = 1$. Além disso, seja a amplitude do tubo horizontal $gg = \frac{1}{100}$ ou $g = \frac{1}{10}$ e seu comprimento $b = 100$ e um lume $kk = \frac{1}{10000}$: será $\mathfrak{f} = 0$ e $\mathfrak{g} = \frac{1}{10000}$, então verdadeiramente

$$\mathfrak{A} = 1 - \frac{1}{40} + \frac{1}{2 \cdot 1600} - \text{etc.}$$

ou

$$\mathfrak{A} = 0{,}97531 \qquad e \qquad \mathfrak{B} = 0{,}77880 \qquad e \qquad A = 98{,}760\,.$$

87. Com estes valores estabelecidos aparecerá a altitude que é devida pela velocidade do efluxo

$$v = 69{,}703 \text{ pés}^{15}.$$

Mas se, no entanto, seja criado um lume no dorso do tubo de cerca de $cc$, que através dele a água irrompa verticalmente, o jorro d'água da fonte se elevará a uma altitude do tamanho de $69\frac{2}{3}$ pés, e por esta razão se afastará com 30 pés da altitude da água no vaso. Verdadeiramente este salto também será diminuído de uma quantidade indefinida por causa da resistência da atmosfera, de modo que a fonte nem mesmo alcance a altitude de $69\frac{2}{3}$ pés.

## CASO IV
## CASO A ÁGUA DO VASO CONSTANTEMENTE CHEIO ESCORRA ATRAVÉS DE UM TUBO CILÍNDRICO INCLINADO

88. Seja (Fig. 8) a amplitude do vaso $AA = ff$ e de altura $AB = a$, o comprimento do tubo inclinado $BC = b$, amplitude $BB = gg$ e o ângulo, que está inclinado com direção à reta vertical, $= \zeta$, de modo que $\cos\zeta$, exprima o seno da inclinação em relação ao horizonte; verdadeiramente eflui água através do lume $CC = kk$, que diante de $ff$ seja mínimo, para que seja

$$\frac{k^4}{f^4} = \mathfrak{f} = 0\,.$$

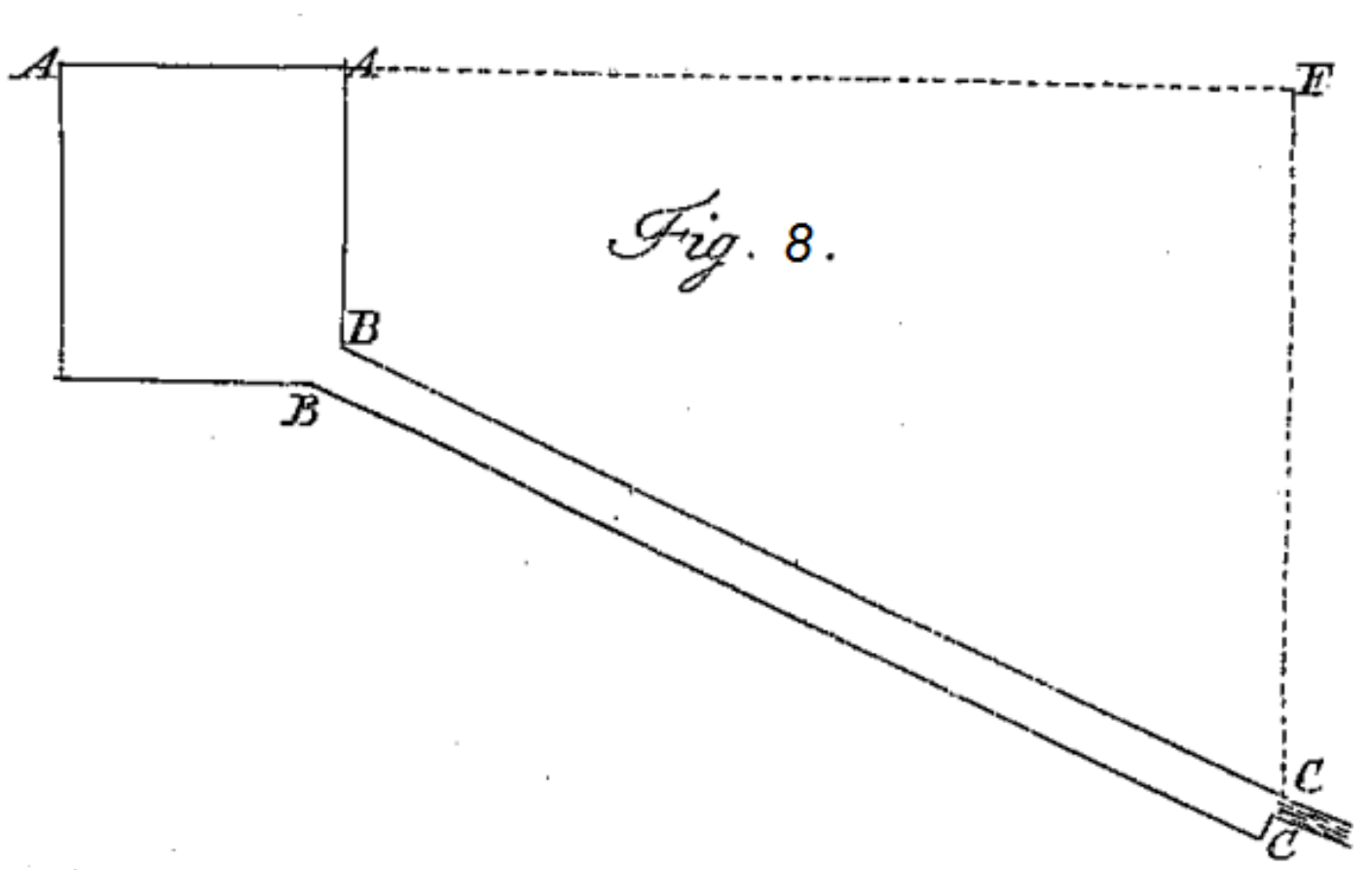

[15] Edição primeira: 69,701. C.T. Corrigiu

Além disso seja colocado

$$\frac{k^4}{g^4} = \mathfrak{g}, \quad \mathfrak{A} = e^{\frac{-\propto a}{f}}, \quad A = \frac{1 - \mathfrak{A}}{\propto}, \quad \mathfrak{B} = e^{\frac{-\propto b}{g}}, \quad \text{e} \quad B = \frac{1 - \mathfrak{B}}{\propto}\;;$$

com estes tendo sido colocados, será a altitude devida pela velocidade, com a qual a água eflui através de $CC$

$$v = \frac{A\mathfrak{B}f + Bg\cos\zeta - (1 - \mathfrak{A}\mathfrak{B})l}{1 {-\propto} \mathfrak{B}\mathfrak{g}}\,.$$

89. Não sendo a altura do vaso $AB = a$ assaz grande e a sua amplitude $ff$ enorme, será $Af = a$ e $\mathfrak{A} = 1$, aproximadamente, donde será obtido

$$v = \frac{a\mathfrak{B} + \frac{1}{\propto}(1 - \mathfrak{B})g\cos\zeta - (1 - \mathfrak{B})l}{1 - (1 - \mathfrak{B})\mathfrak{g}}\,,$$

onde, não podendo $\mathfrak{g}$ superar a unidade e sendo $\mathfrak{B} < 1$, o denominador será uma quantidade positiva; donde o movimento atingirá a uniformidade. Contudo se o numerador fosse ou $= 0$ ou uma quantidade negativa, a água certamente não efluiria, pelo contrário em quietude permanecerá.

90. Se a altura do vaso $AB = a$ seja até certo ponto mínima, o efluxo não está com disposição de se expor, a não ser que seja

$$g\cos\zeta > \propto l$$

ou a não ser que a projeção da declividade do tubo seja maior que $\frac{\propto l}{g}$. Então que seja colocado

$$g = n \propto l = 0{,}0075\, n \text{ pés},$$

para que a água escorra através do tubo $BC$, é necessário que seja

$$\cos\zeta > \frac{1}{n}.$$

Portanto, se $n = 1$ ou se $n < 1$, a água através de tal tubo certamente não escorrerá; razão pela qual, para que a água escorra, é necessário que seja $n > 1$ ou $g > \propto l$, e então a projeção da declividade deve ser maior que $\frac{1}{n}$.

91. Daqui, a declividade do leito pode ser designada em favor das correntes, de tal modo que, caso a declividade fosse menor, a água ficasse estagnada. Dependerá, no entanto, esta declividade da letra $g$, que é declarada como a profundidade da corrente. Então, a fim de que definamos esta declividade para qualquer profundidade de corrente, seja a altitude diante da distância de mil pés, através do qual o leito se estabelece, $=\ z$ pés e será $\frac{z}{1000}$ o seno da declividade. Pelo que, portanto, se a profundidade da corrente seja $g$ pés, por isso $g = \frac{3}{400}n$ e $n = \frac{400}{3}g$ será feito $\frac{z}{1000} > \frac{3}{400}g$: então para que a água escorra através do leito, é necessário que seja

$$z > \frac{30}{4g} \qquad \text{ou} \qquad z > \frac{1000 \propto l}{g}.$$

92. Portanto, poderemos definir a declividade do leito para qualquer profundidade da corrente, que convenha para uma distância de mil pés, no momento em que pela primeira vez a água segue o fluxo: de tal modo que se a declividade fosse menor, a água ficaria estagnada, que gera a tabela seguinte.

| Profundidade da Corrente pé | Declividade diante da dist. 1000 pés pé | Profundidade da Corrente pé | Declividade diante da dist. 1000 pés pé |
|---|---|---|---|
| 0,5 ped. | 15,00 ped. | 5 ped. | 1,50 ped. |
| 1,0 | 7,50 | 6 | 1,25 |
| 1,5 | 5,00 | 7 | 1,07 |
| 2,0 | 3,75 | 8 | 0,94 |
| 2,5 | 3,00 | 9 | 0,83 |
| 3,0 | 2,50 | 10 | 0,75 |
| 3,5 | 2,14 | 11 | 0,68 |
| 4,0 | $1{,}87\frac{1}{2}$ | 12 | $0{,}62\frac{1}{2}$ |
| 4,5 | 1,67 | 13 | 0,58 |
| 5,0 | 1,50 | 14 | $0{,}53\frac{1}{2}$ |
| | | 15 | 0,50 |

93. Se, portanto, a declividade diante de uma dada profundidade fosse maior, que esta tabela indica, a água escoará; e a sua velocidade chegará logo a seguir ao conhecimento, caso fosse colocado $\mathfrak{g} = 1$, donde será feito

$$v = a + \left(e^{\frac{\propto b}{g}} - 1\right)\left(\frac{1}{\propto} g \cos\zeta - l\right).$$

Portanto, sendo mantida a mesma declividade, fica assim patente que contanto que

$$\cos\zeta > \frac{\propto l}{g},$$

a velocidade da corrente será tanto maior, quanto mais longa for a extensão da corrente. Além disso, convém ser observado, o curso da própria corrente ser acelerado, caso seja diminuído o peso da atmosfera.

94. Visto que perante o caso que desenvolvemos (86, 87), no qual a água do vaso com altura de cem pés por uma distância de cem pés através de um tubo horizontal, cuja amplitude $gg = \frac{1}{100}$ pés, era desviada, coloquemos agora que seja a altura deste mesmo vaso a menor possível e que do interior dele a água através do tubo, inclinado com um ângulo semireto em relação ao horizonte, ser conduzida ao mesmo local $C$, e diante de ser

$$gg = \frac{1}{100} \qquad \text{ou} \qquad g = \frac{1}{10}\ \text{pés}\,,$$

será

$$\cos\zeta > \frac{1}{\sqrt{2}} \qquad \text{e} \qquad b = 100\sqrt{2} = 141{,}4\ \text{pés}\,.$$

95. Sendo, pois, $\frac{\propto b}{g} = 0{,}3536$, será $\mathfrak{B} = 0{,}7022$. Estabeleçamos, para aqui, o orifício $kk$ menor possível, e perante uma quantidade mínima $a$ será

$$v = 0{,}2978\left(\frac{400}{\sqrt{2}} - 30\right) = 75{,}299\ \text{pés}^{16}.$$

Que sendo uma quantidade com $5\frac{2}{3}$ pés maior que o caso precedente, segue que a água neste caso ter subido a uma altitude maior, que o caso precedente.

## CASO V

## CASO A ÁGUA DO VASO $AB$ ESCORRA OU ATRAVÉS DE UM TUBO CILÍNDRICO INCLINADO $BE$ OU ATRAVÉS DE UMA INFLEXÃO $bcE$ E DESSE LUGAR IRROMPA ATRAVÉS DE UM TUBO VERTICAL $cD$[17]

96. Seja mínima a altitude $a$ do vaso $AB$, verdadeiramente máxima a amplitude $ff$: então porque aqui estabeleci desenvolver dois casos, seja em ambas os casos a amplitude do tubo que transporta

$$BB = bb = cc = gg\,,$$

A altura do tubinho vertical $CD = c$ e amplitude $cc = hh$ e um lume em $DD = kk$. Então para o primeiro caso, seja o comprimento do tubo $BE = b$ e que sua inclinação em relação à vertical $= \zeta$; será para o outro caso

$$bc = b\cos\zeta \qquad \text{e} \qquad cE = b\sin\zeta\,.$$

[16] Edição primeira: $\mathfrak{B} = 0{,}7021,\ v = 0{,}2979\left(\frac{400}{\sqrt{2}} - 30\right) = 75{,}309.$ C.T. Corrigiu

[17] Nota do Tradutor: embora não mencionada no texto original, acredita-se que este caso tenha como referência a Figura 9 apresentada no Apêndice.

Além disso, seja colocado

$$\frac{k^4}{g^4} = \mathfrak{g} \qquad \text{e} \qquad \frac{k^4}{h^4} = \mathfrak{h}\,.$$

97. Com estes estabelecidos para o primeiro caso teremos:

$$\mathfrak{A} = e^{\frac{-\propto a}{f}}, A = \frac{a}{f}, \mathfrak{B} = e^{\frac{-\propto b}{g}}, B = \frac{1-\mathfrak{B}}{\propto}, \mathfrak{C} = e^{\frac{-\propto c}{h}} \text{ e } C = \frac{1-\mathfrak{C}}{\propto}\;;$$

donde a velocidade do efluxo será originada da altitude $v$, qual seja

$$v = \frac{a\mathfrak{B}\mathfrak{C} + \frac{1}{\propto}(1-\mathfrak{B})\mathfrak{C}g\cos\zeta - \frac{1}{\propto}(1-\mathfrak{C})h - (1-\mathfrak{B}\mathfrak{C})l}{1-(1-\mathfrak{B})\mathfrak{C}\mathfrak{g} - (1-\mathfrak{C})\mathfrak{h}}$$

ou quando substituídos os valores assumidos:

$$v = \frac{e^{\frac{-\propto b}{g}-\frac{\propto c}{h}}a + \frac{1}{\propto}\left(1-e^{\frac{-\propto b}{g}}\right)e^{\frac{-\propto c}{h}}g\cos\zeta - \frac{1}{\propto}\left(1-e^{\frac{-\propto c}{h}}\right)h - \left(1-e^{\frac{-\propto b}{g}-\frac{\propto c}{h}}\right)l}{1-\left(1-e^{\frac{-\propto b}{g}}\right)e^{\frac{-\propto c}{h}}\mathfrak{g} - \left(1-e^{\frac{-\propto c}{h}}\right)\mathfrak{h}}$$

ou se

$$v = \frac{e^{\frac{-\propto b}{g}-\frac{\propto c}{h}}\left(a - \frac{1}{\propto}g\cos\zeta + l\right) + e^{\frac{-\propto c}{h}}\left(\frac{1}{\propto}g\cos\zeta + \frac{1}{\propto}h\right) - \frac{1}{\propto}h - l}{1-\mathfrak{h} - e^{\frac{-\propto c}{h}}(\mathfrak{g}-\mathfrak{h}) + e^{\frac{-\propto b}{g}-\frac{\propto c}{h}}\mathfrak{g}}.$$

98. Verdadeiramente para o outro caso, em vista do exposto no parágrafo 39, teremos:

$$\mathfrak{A} = 1, Af = a,\ \mathfrak{B} = e^{\frac{-\propto b\cos\zeta}{g}},\ \cos\zeta = 1, g = g, \mathfrak{g} = \mathfrak{g} \quad \text{e} \quad B = \frac{1-\mathfrak{B}}{\propto}.$$

Então

$$\mathfrak{C} = e^{\frac{-\propto b\,sin\,\zeta}{h}}, \qquad \cos\eta = 0, \qquad h = g, \qquad \mathfrak{h} = \mathfrak{g} \qquad \text{e} \qquad C = \frac{1-\mathfrak{C}}{\propto}$$

e também

$$\mathfrak{D} = e^{\frac{-\propto c}{h}}, \qquad \cos\theta = -1, \qquad i = h, \qquad \mathfrak{i} = \mathfrak{h} \quad \text{e} \quad D = \frac{1-\mathfrak{D}}{\propto}\ ;$$

com tais valores substituídos, produzirá a altitude que é devida pela velocidade do efluxo

$$v = \frac{(l+a)e^{\frac{-\propto b\cos\zeta}{g}-\frac{\propto b\,sin\,\zeta}{g}-\frac{\propto c}{h}}a + \frac{1}{\propto}\left(1-e^{\frac{-\propto b\cos\zeta}{g}}\right)e^{\frac{\propto b\,sin\,\zeta}{g}}g - \frac{1}{\propto}\left(1-e^{\frac{-\propto c}{h}}\right)h - l}{1-\left(1-e^{\frac{-\propto b\cos\zeta}{g}}\right)e^{\frac{-\propto b\,sin\,\zeta}{g}-\frac{\propto c}{h}}\mathfrak{g} - \left(1-e^{\frac{-\propto b\,sin\,\zeta}{g}}\right)e^{-\frac{\propto c}{h}}\mathfrak{g} - \left(1-e^{\frac{-\propto c}{h}}\right)\mathfrak{h}}$$

ou

$$v = \frac{e^{\frac{-\propto b\cos\zeta-\propto b\,sin\,\zeta}{g}-\frac{\propto c}{h}}\left(a-\frac{1}{\propto}g+l\right) + e^{\frac{-\propto b\,sin\,\zeta}{g}-\frac{\propto c}{h}}\frac{1}{\propto}g + e^{\frac{-\propto c}{h}}\frac{1}{\propto}h - \frac{1}{\propto}h - l}{1-\mathfrak{h}-e^{-\frac{\propto c}{h}}(\mathfrak{g}-\mathfrak{h}) + e^{\frac{-\propto b\cos\zeta}{g}-\frac{\propto b\,sin\,\zeta}{g}-\frac{\propto c}{h}}\mathfrak{g}}.$$

99. Também se deve notar, que embora a altura do tubinho $CD = c$ fosse pequena, entretanto diante da razão anteriormente exposta, esta deve ser tomada maior, e quanto menor fosse a sua amplitude $hh$, tanto mais deva ser aumentada a real altura de $c$. Mesmo se, portanto, o tubo $CD$ fosse o mais estreito ou perfurada uma pequena abertura por uma lâmina, a fração $\frac{\alpha c}{h}$ por isso terá um valor tanto maior, quanto menor fosse a pequena abertura ou quanto menor fosse $h$. Portanto, sendo neste caso $e^{\frac{-\alpha c}{h}}$ um número assaz pequeno, é evidente que a pequena abertura tanto menor pode ser feita, de modo que a água por causa disto absolutamente não eflua.

100. Se o tubinho $DC$ fosse aberto internamente em $DD$, será $\mathfrak{h} = 1$ e para o primeiro caso será obtido

$$v = \frac{e^{\frac{-\propto b}{g}}\left(a+l-\frac{1}{\propto}g\cos\zeta\right) + \frac{1}{\propto}(g\cos\zeta + h) - e^{\frac{\propto c}{h}}\left(l+\frac{1}{\propto}h\right)}{1+\left(e^{\frac{-\propto b}{g}}-1\right)\mathfrak{g}}.$$

Também, diante do segundo caso será obtido

$$v = \frac{e^{\frac{-\propto b\cos\zeta - \propto b\,sin\,\zeta}{g}}\left(a + l - \frac{1}{\propto}g\right) + \frac{1}{\propto}e^{\frac{-\propto b\,sin\,\zeta}{g}}g + \frac{1}{\propto}h - e^{\frac{\propto c}{h}}\left(l + \frac{1}{\propto}h\right)}{1 + \left(e^{\frac{-\propto b\cos\zeta - \propto b\,sin\,\zeta}{g}} - 1\right)\mathfrak{g}};$$

portanto,todas as vezes que for $\mathfrak{g} < 1$, o movimento da água atingirá o estado de uniformidade.

101. O primeiro caso é o que mais serve para ser determinada a retardação da água originada pela fricção através do aqueduto. De cujo movimento que exibimos um exemplo, seja a altitude $a$ da água no receptáculo muito pequena, o comprimento do aqueduto $BE = b =$ 2000 pés, a amplitude $gg = \frac{16}{25}$, ou $g = \frac{4}{5}$ pés, altura $bc$, ou $b\cos\zeta = 150$ pés, também a abertura $kk$ tão pequena, que $\mathfrak{g} = \frac{k^4}{g^4}$ possa ser tomado como zero. Finalmente, seja $c = \frac{1}{10}$ pés, a altura do tubinho $CD$ e $h = \frac{1}{100}$ pés.

102. Com estes estabelecidos, assumido $\alpha = \frac{1}{4000}$ será $\frac{\alpha b}{g} = \frac{5}{8}$, e por esta razão

$$e^{\frac{-\alpha b}{g}} = 0{,}535261^{18} \qquad \text{e} \qquad e^{\frac{\alpha c}{gh}} = 1{,}002503\,.$$

Logo daqui será

$$v = \frac{1}{2}a + 97{,}41988 \text{ pés}^{19}\,.$$

Logo, para este caso, contanto que a altitude do vaso $AB$ fosse de seis pés, a água através do lume $DD$ jorrará para a altitude de 100 pés, assim a fricção tira 50 pés do jato.

103. Se com os remanescentes mantidos os mesmos, seja $gg = 1$ e $g = 1$, será

[18] Edição primeira: 0,535255. C.T. Corrigiu
[19] Edição primeira: 97,42124. C.T. Corrigiu

$$\frac{\alpha b}{g} = \frac{1}{2} \qquad \text{e} \qquad e^{\frac{-\alpha b}{g}} = 0{,}606531\,,$$

e daqui é obtido

$$v = \frac{3}{5}a + 106{,}06 \text{ pés}^{20}\,,$$

de tal maneira que a água haveria de se elevar nove pés mais alto, do que o caso precedente, devido a amplitude aumentada do tubo $BE$. Mas se pelo contrário a amplitude do tubo for diminuída, de tal maneira que seja $gg = \frac{9}{25}$ ou $g = \frac{3}{5}$ pés, diante de $\frac{\alpha b}{g} = \frac{5}{6}$ será

$$e^{\frac{-\alpha b}{g}} = 0{,}43460$$

e por esta razão, a altitude devida pela velocidade do efluxo

$$v = \frac{3}{7}a + 84{,}635 \text{ pés}^{21}$$

e desse modo cai mais de 12 pés da altitude do parágrafo precedente. Todavia se for $gg = \frac{1}{4}$ ou $g = \frac{1}{2}$ pés, de modo que seja obtido $\frac{\alpha b}{g} = 1$, será

$$e^{\frac{-\alpha b}{g}} = 0{,}36788$$

e por esta razão

$$v = \alpha + 75{,}679 \text{ pés}^{22}\,;$$

portanto, agora a altitude do jato é 9 pés menor que anteriormente.

104. Portanto fica patente que nas fontes com jorros d'água a altitude do jato não depende só da altura do receptáculo ou do reservatório, que geralmente os Autores hidráulicos designam, mas também principalmente da amplitude e do comprimento dos canais, através dos quais a água do interior do reservatório é desviada para as fontes de jorros d'água. Pois quanto mais amplos e mais curtos forem os canais, tanto mais próximo a fonte de jato d'água atinge a altitude do reservatório, no entanto, com os canais mais estreitos e com os excessivamente longos que pode se tornar muito estendidos, a tal ponto que a água acenda a uma altitude nula. A causa desta debilitação é, pois, a fricção, aqui expressa pela letra $\alpha$, de cujo valor aqui coloquei

$$\alpha = \frac{1}{4000}\ ;$$

no entanto, considerados certos experimentos, é visto que deve ser colocado

$$\alpha = \frac{1}{4540}\,.$$

[20] Edição primeira: 106,24. C.T. Corrigiu
[21] Edição primeira: 84,810. C.T. Corrigiu
[22] Edição primeira: 75,854. C.T. Corrigiu

## APÊNDICE ACERCA DAS FONTES DE JATOS D'ÁGUA

105. Portanto, daqui se poderá definir cuidadosamente a altitude, para a qual a água nas fontes de jorros d'água será lançada para cima. Com efeito, seja colocada (Fig. 9) primeiramente a altitude da água no reservatório ou $AB = a$ tão pequena, que à vista da própria elevação acima do orifício da fonte $DD$

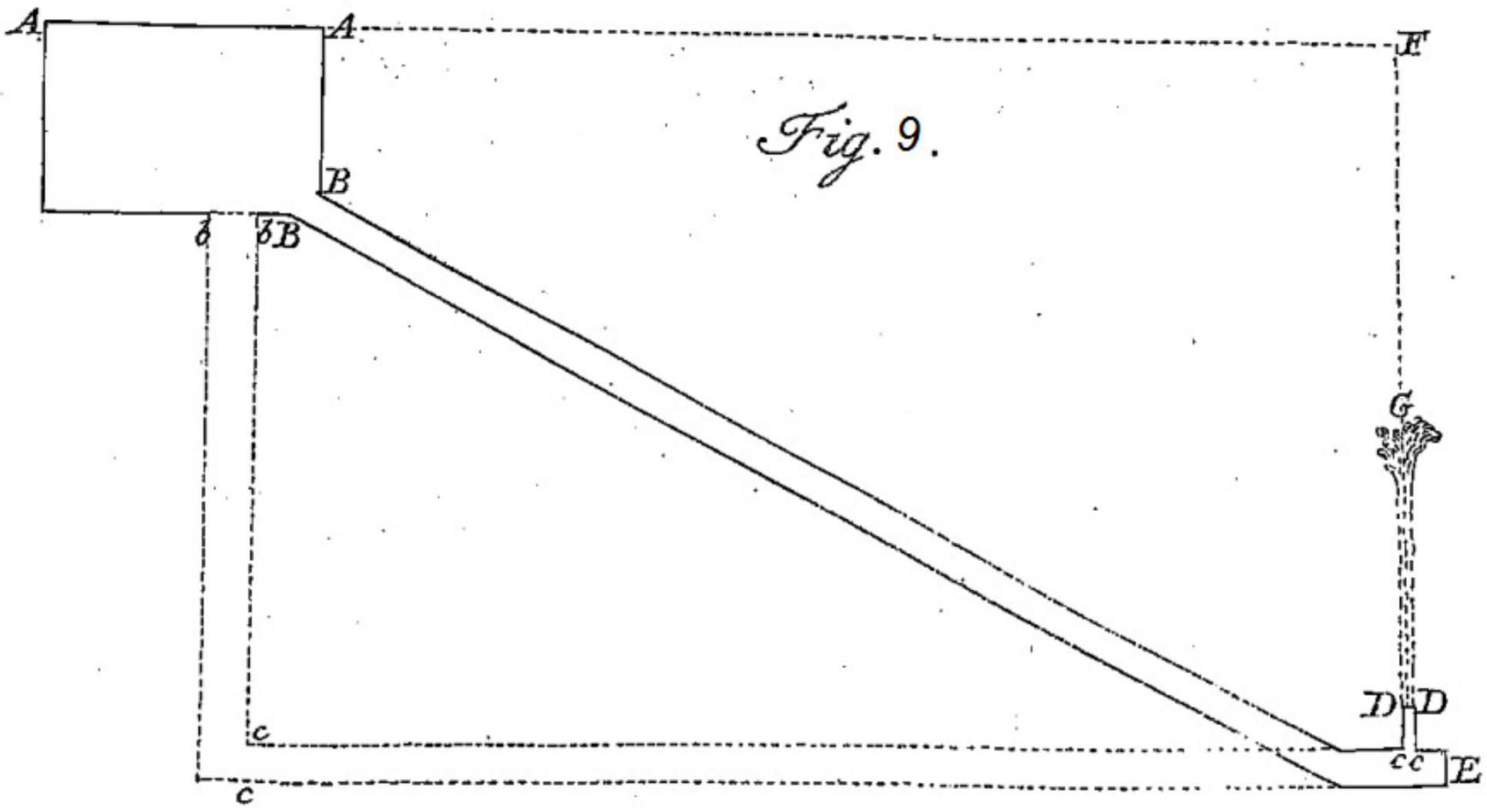

possa ser avaliada como zero. Além disto, verdadeiramente o tubinho $CD$ nem muito longo nem muito estreito, tal que $\frac{\alpha c}{h}$ [23] seja uma fração menor quanto possível; pois para aqui geralmente se submetem todos os casos das fontes de jorros d'água. Com aqueles tendo sido estabelecidos, será a

[23] Na Reprodução de 1954 e na Edição primeira: $\frac{ac}{h}$. Tradutor Corrigiu

altitude devida pela velocidade, com que a água jorrará através do orifício $DD$,

$$v = \frac{g\cos\zeta}{\alpha}\left(1 - e^{\frac{-\alpha b}{g}}\right) - l\left(1 - e^{\frac{-\alpha b}{g}}\right).$$

106. Nesta expressão $l$ denota a altitude da coluna d'água com pressão igual ao peso da atmosfera, e será, portanto, aproximadamente $l = 30$ pés. Além disso, $b$ exprime o comprimento total do aqueduto $BE$, através do qual a água é desviada desde o reservatório até a fonte $DG$; que se colocada disposta segundo uma linha reta, como é geralmente o costume, será $b\cos\zeta$ a altitude da água no reservatório acima da fonte. Por isso, portanto, se esta altitude for declarada $= q$, será $\cos\zeta = \frac{q}{b}$: e deste modo será

$$v = \left(\frac{gq}{\alpha b} - l\right)\left(1 - e^{\frac{-\alpha b}{g}}\right).$$

107. Além disso assumimos todo o aqueduto cilíndrico, assim que deste modo a sua amplitude[24] seja em toda parte a mesma, $= gg$; se, portanto, o diâmetro deste aqueduto seja $= d$, será

$$gg = \frac{1}{4}\pi dd \qquad \text{e} \qquad g = \frac{1}{2}d\sqrt{\pi}\,.$$

Pois, tendo sido coletado através de experimentos que é $\alpha = \frac{1}{4540}$, será

$$\frac{g}{\alpha} = 2270d\sqrt{\pi} = 4023d\,.$$

Portanto, um cálculo satisfatoriamente preciso se conduzirá, caso assumamos para o diâmetro do canal $= d$

$$\frac{g}{\alpha} = 4000d\,.$$

108. Então sendo

$$v = \frac{g}{\alpha b}\left(q - \frac{\alpha b l}{g}\right)\left(1 - e^{\frac{-\alpha b}{g}}\right),$$

[24] Edição primeira: altitude. C.T. Corrigiu

diante de

$$1-e^{\frac{-\alpha b}{g}}=\frac{\propto b}{g}-\frac{\propto^2 b^2}{2g^2}+\frac{\propto^3 b^3}{6g^3}-\frac{\propto^4 b^4}{24g^4}+etc.$$

será obtida a altitude do jato vertical:

$$v=\left(q-\frac{\alpha bl}{g}\right)\left(1-\frac{\propto b}{2g}+\frac{\propto^2 b^2}{6g^2}-\frac{\propto^3 b^3}{24g^3}+etc.\right)\ ;$$

onde é

$$\frac{\propto b}{g}=\frac{b}{4000d}\,.$$

109. Portanto, eis que segue a regra descoberta sobre a altitude do jato em qualquer fonte de jato d'água: *Seja dividido o comprimento total do canal pelo diâmetro da amplitude do mesmo canal e o resultado seja colocado* $= n$*, então será*

$$N=1-\frac{n}{2\cdot 4000}+\frac{n^2}{6\cdot 4000^2}-\frac{n^3}{24\cdot 4000^3}+\frac{n^4}{120\cdot 4000^4}-etc.$$

*Além disto* $q$ será *a elevação da água no reservatório acima da fonte, e será a altitude do jato*

$$v=Nq-\frac{3}{400}Nn\text{ pés.}$$

*Ou colocando* $\frac{3}{400}Nn=M$ *será a altitude do jato expressa em pés*

$$v=Nq-M\,.$$

110. Portanto, a partir de um dado numero $n$, que revela, um certo comprimento do canal dividido pelo seu diâmetro, sejam coletados os valores das letras $N$ e $M$; então primeiro seja multiplicado o número $N$ pela elevação da água no reservatório acima da fonte, cuja altitude tenha sido dada em pés, e deste produto seja subtraído o segundo número, $M$, e o resíduo revelará a altitude do jato expressa em pés.

111. No entanto, este modo não somente produzirá a própria altitude do jato, como também a altura devida pela velocidade, com que a

água jorra; efetivamente diante desta altitude subsiste a resistência do ar que adicionalmente a diminui um pouco, de tal maneira que, se a fonte jorrar verticalmente, a altitude será um pouco menor, que aquela obtida por meio da regra. Verdadeiramente deste abatimento, a qual tem a origem em outra causa, aqui não haverá nenhuma consideração.

112. Portanto, para qualquer caso apresentado, se possa obter facilmente a altitude do jato diminuída somente devido à fricção, convirá ser construída uma tabela, a qual para qualquer valor do próprio $n$, ou descoberto qual o número da divisão do comprimento do canal pelo seu diâmetro, forneça os valores das letras $N$ e $M$. Efetivamente, com estes descobertos, se a altura do reservatório acima do orifício da fonte $q$ for introduzida no cálculo, sem dificuldade a altitude do jato $v$ daí será obtida, qual seja

$$v = Nq - M \text{ pés.}$$

TABELA[25]

que apresenta os valores das letras $N$ e $M$ para cada um dos valores da letra $n$

| $n$ | $N$ | $M$ |
|---|---|---|
| 100 | 0,9876 | 0,7407 |
| 200 | 0,9754 (+ 2) | 1,4631 (+ 3) |
| 300 | 0,9634 (+ 1) | 2,1677 (+ 1) |
| 400 | 0,9516 | 2,8549 (— 1) |
| 500 | 0,9400 | 3,5251 (— 1) |
| 600 | 0,9286 | 4,1788 (— 1) |
| 700 | 0,9174 | 4,8163 (— 1) |
| 800 | 0,9063 (+ 1) | 5,4381 |
| 900 | 0,8955 | 6,0445 (— 1) |
| 1 000 | 0,8848 | 6,6359 (— 2) |
| 1 100 | 0,8743 | 7,2128 (+ 1) |
| 1 200 | 0,8639 | 7,7755 (— 1) |
| 1 300 | 0,8538 (— 1) | 8,3242 (— 3) |
| 1 400 | 0,8437 | 8,8594 (— 4) |
| 1 500 | 0,8339 | 9,3813 (— 3) |
| 1 600 | 0,8242 | 9,8904 (— 3) |
| 1 700 | 0,8147 (— 1) | 10,3869 (— 3) |
| 1 800 | 0,8053 (— 1) | 10,8712 (— 4) |
| 1 900 | 0,7960 | 11,3434 (— 3) |
| 2 000 | 0,7869 | 11,8041 (— 3) |
| 2 100 | 0,7780 | 12,2533 (— 3) |
| 2 200 | 0,7692 | 12,6915 (— 3) |
| 2 300 | 0,7605 | 13,1189 (— 4) |

[25] Na tabela de Euler existem muitos erros. Nesta nova tabela, forneci o que foi calculado por R. Weillero. São anexados aos valores da primeira edição pequenos valores entre parêntesis, assim, por exemplo, 8,8594 (-4) signifique que a tabela de Euler exibe o número 8,8590. C.T.

| $n$ | $N$ | $M$ |
| --- | --- | --- |
| 2 400 | 0,7520 (— 1) | 13,5357 (— 3) |
| 2 500 | 0,7436 (— 1) | 13,9422 (— 3) |
| 2 600 | 0,7353 | 14,3386 (— 1) |
| 2 700 | 0,7272 (— 1) | 14,7253 (— 3) |
| 2 800 | 0,7192 (— 1) | 15,1024 (— 4) |
| 2 900 | 0,7113 (— 1) | 15,4703 (— 5) |
| 3 000 | 0,7035 | 15,8290 (— 4) |
| 3 100 | 0,6959 (— 1) | 16,1789 (— 5) |
| 3 200 | 0,6883 | 16,5201 (— 3) |
| 3 300 | 0,6809 | 16,8530 (— 5) |
| 3 400 | 0,6736 (+ 2) | 17,1776 (— 5) |
| 3 500 | 0,6664 | 17,4941 (— 5) |
| 3 600 | 0,6594 (— 1) | 17,8029 (— 3) |
| 3 700 | 0,6524 (— 1) | 18,1041 (— 5) |
| 3 800 | 0,6455 | 18,3978 (— 3) |
| 3 900 | 0,6388 (— 1) | 18,6842 (— 5) |
| 4 000 | 0,6321 | 18,9636 (— 3) |
| 4 100 | 0,6256 (— 1) | 19,2361 (— 5) |
| 4 200 | 0,6191 | 19,5019 (— 4) |
| 4 300 | 0,6127 | 19,7611 (— 6) |
| 4 400 | 0,6065 | 20,0139 (— 3) |
| 4 500 | 0,6003 | 20,2604 (— 4) |
| 4 600 | 0,5942 | 20,5009 (— 4) |
| 4 700 | 0,5882 | 20,7354 (— 3) |
| 4 800 | 0,5823 | 20,9642 (— 2) |
| 4 900 | 0,5765 | 21,1873 (— 4) |
| 5 000 | 0,5708 | 21,4049 (— 5) |
| 5 100 | 0,5652 (— 1) | 21,6171 (— 7) |

| $n$ | $N$ | $M$ |
|---|---|---|
| 5 200 | 0,5596 (— 1) | 21,8240 (— 6) |
| 5 300 | 0,5541 (— 1) | 22,0259 (— 5) |
| 5 400 | 0,5487 (— 1) | 22,2228 (— 4) |
| 5 500 | 0,5434 | 22,4148 (— 3) |
| 5 600 | 0,5381 (+ 1) | 22,6021 (— 4) |
| 5 700 | 0,5330 | 22,7847 (— 3) |
| 5 800 | 0,5279 | 22,9629 (— 2) |
| 5 900 | 0,5229 | 23,1366 |
| 6 000 | 0,5179 | 23,3061 (— 3) |
| 6 100 | 0,5130 | 23,4714 (— 5) |
| 6 200 | 0,5082 | 23,6326 (— 5) |
| 6 300 | 0,5035 | 23,7898 (— 4) |
| 6 400 | 0,4988 (+ 1) | 23,9431 (— 3) |
| 6 500 | 0,4942 | 24,0926 (— 2) |
| 6 600 | 0,4897 | 24,2385 (— 4) |
| 6 700 | 0,4852 | 24,3808 (— 4) |
| 6 800 | 0,4808 | 24,5195 (— 3) |
| 6 900 | 0,4764 (+ 1) | 24,6548 (+ 6) |
| 7 000 | 0,4721 | 24,7868 (— 2) |
| 7 100 | 0,4679 | 24,9155 (— 4) |
| 7 200 | 0,4637 (+ 1) | 25,0410 (— 4) |
| 7 300 | 0,4596 (+ 1) | 25,1635 (— 4) |
| 7 400 | 0,4555 (+ 1) | 25,2829 |
| 7 500 | 0,4515 | 25,3994 (— 5) |
| 7 600 | 0,4476 (— 1) | 25,5129 (— 4) |
| 7 700 | 0,4437 (— 1) | 25,6237 (— 3) |
| 7 800 | 0,4399 (— 1) | 25,7318 (— 2) |
| 7 900 | 0,4361 (— 1) | 25,8372 (— 1) |

| $n$ | $N$ | $M$ |
|---|---|---|
| 8 000 | 0,4323 | 25,9399 (— 1) |
| 8 100 | 0,4286 | 26,0402 (— 3) |
| 8 200 | 0,4250 | 26,1380 (— 3) |
| 8 300 | 0,4214 | 26,2333 (— 2) |
| 8 400 | 0,4179 | 26,3263 (— 2) |
| 8 500 | 0,4144 | 26,4170 (— 2) |
| 8 600 | 0,4109 | 26,5055 (— 3) |
| 8 700 | 0,4075 | 26,5918 (— 3) |
| 8 800 | 0,4042 (— 1) | 26,6759 (— 2) |
| 8 900 | 0,4009 (— 1) | 26,7580 (— 2) |
| 9 000 | 0,3976 | 26,8380 (— 3) |
| 9 100 | 0,3944 | 26,9161 (— 5) |
| 9 200 | 0,3912 (+ 1) | 26,9922 (— 5) |
| 9 300 | 0,3881 (+ 1) | 27,0665 (— 5) |
| 9 400 | 0,3849 (+ 1) | 27,1389 (— 4) |
| 9 500 | 0,3819 | 27,2096 (— 2) |
| 9 600 | 0,3789 | 27,2785 (— 3) |
| 9 700 | 0,3759 (+ 1) | 27,3457 (— 3) |
| 9 800 | 0,3729 (+ 3) | 27,4112 (— 2) |
| 9 900 | 0,3700 (+ 2) | 27,4751 (— 1) |
| 10 000 | 0,3672 | 27,5375 (— 2) |
| 11 000 | 0,3404 | 28,0822 (— 1) |
| 12 000 | 0,3167 | 28,5064 (— 1) |
| 13 000 | 0,2958 | 28,8368 (— 5) |
| 14 000 | 0,2771 | 29,0941 (— 1) |
| 15 000 | 0,2604 | 29,2945 (— 1) |
| 16 000 | 0,2454 | 29,4505 (–101) |
| 17 000 | 0,2319 | 29,5721 (+ 7) |

| $n$ | $N$ | $M$ |
|---|---|---|
| 18 000 | 0,2198 (— 1) | 29,6667 (— 3) |
| 19 000 | 0,2087 (— 1) | 29,7404 (+ 1) |
| 20 000 | 0,1987 (— 1) | 29,7979 (— 1) |
| 21 000 | 0,1895 | 29,8426 (— 1) |
| 22 000 | 0,1811 | 29,8774 (— 1) |
| 23 000 | 0,1733 (+ 1) | 29,9045 (+ 1) |
| 24 000 | 0,1663 | 29,9256 |
| 25 000 | 0,1597 | 29,9421 |
| 26 000 | 0,1536 | 29,9549 (— 2) |
| 27 000 | 0,1480 (— 1) | 29,9649 |
| 28 000 | 0,1427 | 29,9726 (+ 1) |
| 29 000 | 0,1378 | 29,9787 |
| 30 000 | 0,1333 (— 1) | 29,9834 (+ 4) |
| 31 000 | 0,1290 | 29,9871 (+ 3) |
| 32 000 | 0,1250 | 29,9899 (+ 2) |
| 33 000 | 0,1212 | 29,9922 (+ 3) |
| 34 000 | 0,1176 | 29,9939 (+34) |
| 35 000 | 0,1143 | 29,9952 |
| 36 000 | 0,1111 | 29,9963 (+ 1) |
| 37 000 | 0,1081 | 29,9971 (—34) |
| 38 000 | 0,1053 | 29,9978 (+ 1) |
| 39 000 | 0,1026 (— 1) | 29,9983 |
| 40 000 | 0,1000 | 29,9986 (+ 1) |
| 41 000 | 0,0976 | 29,9989 (+ 1) |
| 42 000 | 0,0952 | 29,9992 |
| 43 000 | 0,0930 | 29,9994 |
| 44 000 | 0,0909 | 29,9995 |
| 45 000 | 0,0889 | 29,9996 |

| $n$ | $N$ | $M$ |
|---|---|---|
| 50 000 | 0,0800 | 29,9999 |
| 55 000 | 0,0727 | 30,0000 |
| 60 000 | 0,0667 (— 1) | 30,0000 |
| 65 000 | 0,0615 | 30,0000 |
| 70 000 | 0,0571 | 30,0000 |
| 75 000 | 0,0533 | 30,0000 |
| 80 000 | 0,0500 | 30,0000 |
| 85 000 | 0,0471 (— 1) | 30,0000 |
| 90 000 | 0,0444 | 30,0000 |
| 95 000 | 0,0421 | 30,0000 |
| 100 000 | 0,0400 | 30,0000 |
| 110 000 | 0,0364 | 30,0000 |
| 120 000 | 0,0333 | 30,0000 |
| 130 000 | 0,0308 | 30,0000 |
| 140 000 | 0,0286 | 30,0000 |
| 150 000 | 0,0267 | 30,0000 |
| 160 000 | 0,0250 | 30,0000 |
| 170 000 | 0,0235 | 30,0000 |
| 180 000 | 0,0222 | 30,0000 |
| 190 000 | 0,0211 | 30,0000 |
| 200 000 | 0,0200 | 30,0000 |
| 300 000 | 0,0133 | 30,0000 |

113. Não é tarefa, que esta tabela seja continuada mais além; se, com efeito, $n$ for um número maior que 50 000, será justamente

$$N = \frac{4000}{n} \qquad \text{e} \qquad M = 30 \text{ pés}.$$

Logo, qualquer que fosse a distância ou o comprimento do canal e do diâmetro, a partir de $n$ calculado, que surge da divisão do comprimento pelo diâmetro, com a ajuda desta tabela facilmente são selecionados os valores $N$ e $M$, que tendo sido descobertos, se a elevação da água no reservatório acima da fonte seja colocada $= q$, será a altura da fonte

$$v = Nq - M \text{ pés}\,.$$

114. Se nula fosse a fricção, a altura da fonte $v$ seria igual à altitude do reservatório $q$ ou $q = v$. Donde fica exposto que por causa da fricção esta altitude $q$ diminui com duplo modo: com efeito, primeiro deve ser multiplicada por $N$, que é um número menor que a unidade; então verdadeiramente, além disto, deste produto se deve subtrair a altura $M$, a qual não pode superar 30 pés, e ainda esta última diminuição é devida à pressão da atmosfera, a qual se maior ou menor fosse, os números $M$ pela mesma razão devessem aumentar ou diminuir.

115. Portanto, daqui fica claro, que se a altura do reservatório $q$ for menor que $\frac{M}{N}$, a água não sai do orifício $DD$, pois o movimento dela é profundamente restringido pela fricção. Desse modo se fosse $n = 100\,000$, a fonte não dará jato, a não ser que seja

$$q > \frac{3000}{4} \qquad \text{ou} \qquad q > 750 \text{ pés}\,.$$

No entanto, por outro lado, se a altura $q$ fosse concedida, para que a água pelo menos eflua, deve ser $\frac{M}{N} < q$; pois como cresça o valor $\frac{M}{N}$ com o aumento de $n$, logo chegará ao conhecimento o limite, abaixo do qual deve cessar o valor do próprio $n$.

116. Se o local do reservatório for dado juntamente com o local da fonte e seja procurado este duto de água, com o qual a fonte acomete a máxima altitude, primeiro deve ser dirigido o duto do reservatório segundo uma linha reta o quanto for possível; depois, verdadeiramente desejável de preferência que se fique satisfeito, se foi atribuída a máxima amplitude ao canal, quanto as circunstâncias permitam. Com efeito, da tabela a altitude do jato mostra-se antes de tudo depender da amplitude do canal.

117. Afim de que seja examinado um exemplo, seja a altitude do reservatório $q = 130$ pés e a distância dele até a fonte $b = 2500$ pés e

diante dos diâmetros de vários canais a altitude do jato deste modo se conhecerá:

| Diâm. do canal | $1$ | $\frac{1}{2}$ | $\frac{1}{3}$ | $\frac{1}{4}$ | $\frac{1}{5}$ | $\frac{1}{6}$ |
|---|---|---|---|---|---|---|
| número $n$ | 2500 | 5000 | 7500 | 10 000 | 12 500 | 15 000 |
| $N$....... | 0,7436 | 0,5708 | 0,4515 | 0,3672 | 0,3059[26] | 0,2604 |
| $M$........ | 13,942 | 21,405 | 25,399 | 27,538 | 28,682[26] | 29,295 |
| alt. do jato $v$ | 82,72 | 52,80 | 33,30 | 20,19 | 11,10[26] | 4,56 |

Portanto, se o diâmetro do canal fosse menor que $\frac{1}{6}$ pé, a água claramente não saltará.

118. Além disso, fica evidente que a água não pode ser trazida de local muito afastado, a não ser que esse local seja bastante elevado: por essa razão, não é permitido que os canais sejam feitos demasiadamente amplos. Desse modo, se a água for retirada de uma distância de um milhar, tal como seja $b = 25\ 000$, e o diâmetro dos canais seja de $\frac{1}{4}$ pé, será $n = 100\ 000$, que não pode ser colocada à disposição, a menos que o reservatório esteja elevado a mais de 750 pés acima do local da fonte. Portanto, estabeleçamos que esta elevação seja $q = 1000$ pés e a água da fonte não ascenderá a uma altitude maior que 10 pés: mas se fosse $q = 2000$ pés, a altura do jato avançasse a 50 pés e para cada mil pés, com os quais a altura $q$ é aumentada, a altitude do salto aumenta em 40 pés de cada vez.

119. Se a atmosfera exercesse pressão nula, devesse ser omitida a quantidade $M$; e neste caso a água irrompesse através do orifício $DD$, como se a água no reservatório fosse mais elevada. Mas fosse $v = Nq$, ou se diante da fricção se tivesse $CG$ junto de $CF$, como que o número $N$ próximo da unidade; todavia é $N < 1$, a menos que $n = 0$, e os principais valores do próprio $N$ para sequências de valores do próprio $n$ corresponderão:

| $N = 1$ | $\frac{3}{4}$ | $\frac{2}{3}$ | $\frac{1}{2}$ | $\frac{1}{3}$ | $\frac{1}{4}$ | $\frac{1}{5}$ | $\frac{1}{6}$ | $\frac{1}{7}$ | $\frac{1}{8}$ | $\frac{1}{9}$ |
|---|---|---|---|---|---|---|---|---|---|---|
| $n = 0$ | 2400 | 3500 | 6300 | 11300 | 15700 | 19860 | 24000 | 28000 | 32000 | 36000 |

[26] Edição primeira: 0,3062, 28,660, 11,19. C.T. Corrigiu

Daqui se fosse $N = \frac{1}{\nu}$ e $\nu > 5$, será $n = 4000\,\nu$.

120. No entanto, diante da pressão da atmosfera, a altura da fonte $CG$ é mais reduzida e da altura já diminuída $Nq$ se deve, além disso, subtrair a altura $M$, de tal modo que apareça:

$$v = Nq - M\,.$$

Também, fica patente da tabela anexa, que todas as vezes que o número $\frac{b}{d} = n$ for maior que 50 000, será

$$v = \frac{4000d}{b}q - 30\ \text{pés}\,.$$

Pois sendo a pressão da atmosfera variável, e proporcional com a altura do mercúrio no barômetro, segue que as fontes de jato d’água devem por isso ascender para uma altura maior, quanto menos elevado for o mercúrio no barômetro.